\def\degree{\hbox{$^\circ$}}

\documentclass[preprint,12pt,authoryear]{elsarticle}

\usepackage{amssymb}
\usepackage{natbib}
\usepackage{graphicx}
\usepackage{hyperref}

\journal{Earth and Planetary Science Letters}

\begin{document}

\begin{frontmatter}



\title{The diurnal cycle and temporal trends of surface winds}


\author[BIDR]{Yosef Ashkenazy\corref{cor1}}
\ead{ashkena@bgu.ac.il}
\cortext[cor1]{corresponding author}
\author[BIDR]{Hezi Yizhaq}
\address[BIDR]{Department of Solar Energy and Environmental Physics, BIDR,
Ben-Gurion University, Midreshet Ben-Gurion, Israel}

\begin{abstract}
  Winds play an essential role in the climate system.  In this study, we analyze the global pattern of the diurnal cycle of surface (10 m) winds from the ERA5 reanalysis data. We find that over the land and especially over sand dune regions, the maximal wind speed and wind drift potential (DP) occur during the hours around midday. However, over the ocean, the wind also peaks at night. Using the sensible heat flux, we show that the weaker winds over land at night are due to a nocturnal cooling that decouples upper atmospheric levels and their associated stronger winds from the surface---nocturnal cooling is much smaller over the ocean. We also analyze wind data from more than 400 meteorological stations in the USA and find a similar diurnal trend as in the reanalysis data. The timing (during the day) of the maximum wind speed has not varied much over the past 70 years. Yet, the wind speed, wind power, and wind drift potential exhibit significant increases with time over the ocean and, to a much lesser degree, over the land and sand dune regions. We compare the USA and Europe DP and wind speed of the ERA5 to that of meteorological stations and find that the ERA5 significantly underestimates real winds; however, the temporal patterns of the two are similar.
\end{abstract}


% \begin{highlights}
% \end{highlights}


\end{frontmatter}

%\linenumbers

\section{Introduction}
\label{sec:intro}

Winds are generated, in part, due to pressure gradients associated with spatial and temporal temperature variations. They influence our daily life and are a major component of the climate system on a wide range of temporal and spatial scales. Winds drive the ocean and are a major source of oceanic energy~\cite[][]{Munk-1966:abyssal, Munk-Wunsch-1998:abyssal, Wunsch-2020:is}. Extreme winds (as in hurricanes) are a major threat to human life and livelihood, causing substantial destruction. Thus, it is not surprising that many studies have investigated different aspects of wind dynamics and characteristics.

Winds are also the driving force of sand dunes~\cite[][]{Bagnold-1941:physics} and dust emission and transport. Dunes cover a substantial area of the world's deserts and form a unique physical and ecological system~\cite[][]{Pye-1982:Morpholgical, Veste-Littmann-Breckle-Yair-2001:role, Shanas-Abu-Galyun-Alshamlih-et-al-2006:reptile, Pye-Tsoar-2008:aeolian, Thomas-Wiggs-2008:aeolian, Tsoar-2008:land}. For example, active and stable dunes host various flora species~\cite[][]{Danin-1991:plant, Danin-1996:plants, Maun-2009:biology, Bel-Ashkenazy-2014:effects}, while some of these are endemic species~\cite[][]{Kutiel-2001:conservation, Rocha-Queiroz-Pirani-2004:plant}. While many studies have concentrated on the role played by wind power and direction in dune activity and type (shape)~\cite[][]{Fryberger-1979:dune, Pye-Tsoar-2008:aeolian}, only a few have investigated the diurnal  cycle~\cite[][]{Yang-He-Mamtimin-et-al-2013:diurnal, Zhou-Mamtimin-Pan-et-al-2019:relationship, Gunn-Wanker-Lancaster-et-al-2021:circadian} of the winds and their impact on dune activity. More specifically \cite{Yang-He-Mamtimin-et-al-2013:diurnal} studied 2-years saltation activity in the Taklimakan desert of China and found increased salt saltation during the middle of the day and attributed this increased activity mainly to stronger winds. \cite{Zhou-Mamtimin-Pan-et-al-2019:relationship} found a correlation between the air temperature and sand activity in the same region. \cite{Gunn-Wanker-Lancaster-et-al-2021:circadian} reported increased sand and dust activity during daytime at the White Sands National Park (New Mexico, USA) and associated this activity with stronger winds and drier conditions during the day. Concerning trends in wind activity, a previous study~\cite[][]{Young-Zieger-Babanin-2011:global} analyzed 23 years of oceanic altimetry data and found a general increase in wind speed and wave height. These findings were confirmed in a later study~\cite[][]{Young-Ribal-2019:multiplatform} using satellite data that span a longer period, from 1983 to 2010. Increased wind speed trend over the ocean was also found based on several reanalysis data~\cite[][]{Wohland-Omrani-Witthaut-et-al-2019:inconsistent}; the increasing trend of wind speed was less conclusive overland~\cite[e.g.,][]{Zeng-Ziegler-Searchinger-et-al-2019:reversal}. Another study which was based on the ERA-Interim reanalysis~\cite[][]{Yu-Zhong-Sun-2020:climatology} reported an increasing trend in wind speed over Antarctica and the Southern Ocean. The above studies did not focus on sand dune regions and did not use the most recent and detailed reanalysis data, ERA5, in their studies. Here we aim to explore both the diurnal variation of surface winds and the trends in wind activity since 1950; we mainly focus on surface winds over sand dune areas. We do so based on high resolution, relatively recent reanalysis data, the ERA5~\cite[][]{Hersbach-Bell-Berrisford-et-al-2020:era5}.

In brief, we find that over land, the winds are much stronger during the day than at night. Based on previous studies \cite[e.g.,][]{Zhang-Zheng-2004:diurnal, Gunn-Wanker-Lancaster-et-al-2021:circadian}, we attribute these stronger daytime winds to the small heat capacity (small thermal inertia) of the land (i.e., due to the fact that only the upper few cms of the soil are drastically affected by diurnal heating) and the temperature inversion at night, which decouples the upper and active atmospheric levels from the ground. Over the ocean, the heat capacity (thermal inertia) is much larger (i.e., due to the upper several tens of meters of the mixed layer, which are hardly affected by diurnal heating), and the situation is more complex. In some locations, there are two maxima during the 24-hour period, and their timing varies from one latitude to another. As for the trends in wind activity, we find a general increase with time, in which the increase is pronounced over the ocean and weak overland, especially over sand dune regions.

The paper is organized as follows. We first describe the data (Sec.~\ref{sec2_data}) and methods (Sec.~\ref{sec2_methods}). We then present in Sec.~\ref{sec3_1} the results regarding the timing of the maximum wind activity (as reflected by the wind drift potential, wind speed, and wind power) and the results regarding the trend in wind activity as reflected by the annual value of the above measures (Sec.~\ref{sec3_2}). We then summarize the results and draw some conclusions. 

\section{Data and Methods}
\label{sec:data_meth0ds}

\subsection{Data.}\label{sec2_data}
The main results of this study are based on the ERA5 reanalysis~\cite[][]{Hersbach-Bell-Berrisford-et-al-2020:era5},
% (\href{https://cds.climate.copernicus.eu/cdsapp#!/home}{https://cds.climate.copernicus.eu/cdsapp#!/home}),
the fifth-generation atmospheric reanalyses of the European Centre for Medium-Range Weather Forecasts (ECMWF). The spatial resolution of the ERA5 data is 0.25\degree{}, and its temporal resolution is one hour. The ERA5 first spanned the time period from 1979 to the present day and then also included data from 1950 to 1978; we analyzed the entire period from the beginning of 1950 to the end of 2021, for a total of 72 years. Among the many variables and measurands of the ERA5, we focus on the surface (10 m) winds and sensible heat flux. To our knowledge, the ERA5 is the only reanalysis that provides hourly data, which are essential in analyzing the diurnal cycles of the wind. In addition, it provides high spatial resolution (0.25\degree{}) global data in comparison to other global reanalysis sources. 

In addition to the analysis of the ERA5 surface winds, we also analyzed the wind data of 406 meteorological stations in the USA, provided by the National Climatic Data Center (NCDC); see \href{www.ncdc.noaa.gov}{www.ncdc.noaa.gov}. The temporal resolution of these stations is 20 minutes; most of the stations span the period from 2009 to 2018. We also analyze global wind data of the NCDC, from 1973 to 2021. We focus only on stations that had more than 80\% of valid measurements. The number of analyzed stations (generally) increased with time, from 1281 in 1973 to 6718 in 2021, with variability over the years. 

\subsection{Methods. }\label{sec2_methods}

The ERA5 provides the surface (10 m) zonal, $u$, and meridional, $v$, components of the wind, and based on this, we calculated the wind speed, $U=\sqrt{u^2+v^2}$. Then we calculated $U^3$ (which is proportional to the wind power, $\frac{1}{2}A\rho U^3$, where $A$ is the cross-section area and $\rho$ is the air density). In addition, we calculated the instantaneous (hourly) sand drift potential (DP), which is a measure for sand transport. It is defined as
\begin{equation}
  DP=\langle U^2(U-U_{\mathrm{th}}) \rangle
\label{eq:dp}
\end{equation}
where $U$ is the 10-m-height wind speed, and $U_{\mathrm{th}}$ is the minimum wind speed required for sand transport where typically $U_{\mathrm{th}}=12$ knots~\cite[][]{Fryberger-1979:dune}. $\langle \cdot \rangle$ represents an average over time (the sum of elements over the number of elements).

For each month, we calculated the diurnal cycle of the above three measures, $U$, $U^3$, and DP, i.e., we averaged the same hour within the day over all days of the month. The time within the day of the ERA5 is the Greenwich Mean Time (GMT), and we adjusted the time (in hours) at a longitude according to $t=t_g+\frac{24}{360} x$ (i.e., one rotation over 24 hours) where $t$ and $t_g$ is the local time and GMT, respectively, and $-180\le x \le 180$ is the longitude.

We also used the instantaneous sensible heat flux to study the vertical stability of the air close to the surface.  The sensible heat flux is proportional to surface wind speed and the temperature difference between the air and the surface~\cite[][]{Gill-1982:atmosphere}. We note that other air stability measures can be used to explain the diurnal cycle of surface winds; yet, since we are dealing with surface winds, we study the sensible heat flux that is proportional to the temperature difference close to the surface. At each grid point, we calculated the annual mean of the difference between the daily maximum and minimum sensible heat flux and found the annual mean of this measure. We also found the time (within the day) of this measure.

We analyzed the meteorological (NCDC) station data to provide some support for the ERA5 analysis. We used two subsets of the NCDC data: one that includes 406 stations from the USA with 20-minute resolution. Here we simply constructed, for each station, the probability density function (pdf) of the time of maximum speed within the day. We then calculated the mean $\pm$ std of the pdfs of the different stations. Another much larger subset includes stations from the entire world, with a 1--hour (or less) temporal resolution. We used the data quality classification of the NCDC to filter out problematic data points and considered time series that have more than 80\% valid data; see \href{https://www.ncei.noaa.gov/data/global-hourly/doc/isd-format-document.pdf}{https://www.ncei.noaa.gov/data/global-hourly/doc/isd-format-document.pdf}. We considered data from 1973 to 2021, with an increasing number of stations with time (from 1281 in 1973 to 6718 in 2021). Most of the stations are from the USA, Europe, Japan, and southeast Australia. We gridded the data of the different stations (falling on the same grid point) using Gaussian smoothing---see Sec. 3.1 of~\cite{Ashkenazy-Yizhaq-Tsoar-2012:sand} for more details. The spatial resolution of the gridded data is 1\degree{}. 

\section{Results}
\label{sec:results}

\subsection{The timing of the maximum wind.}\label{sec3_1}

\begin{figure}
  \begin{center}
  \includegraphics[width=30pc]{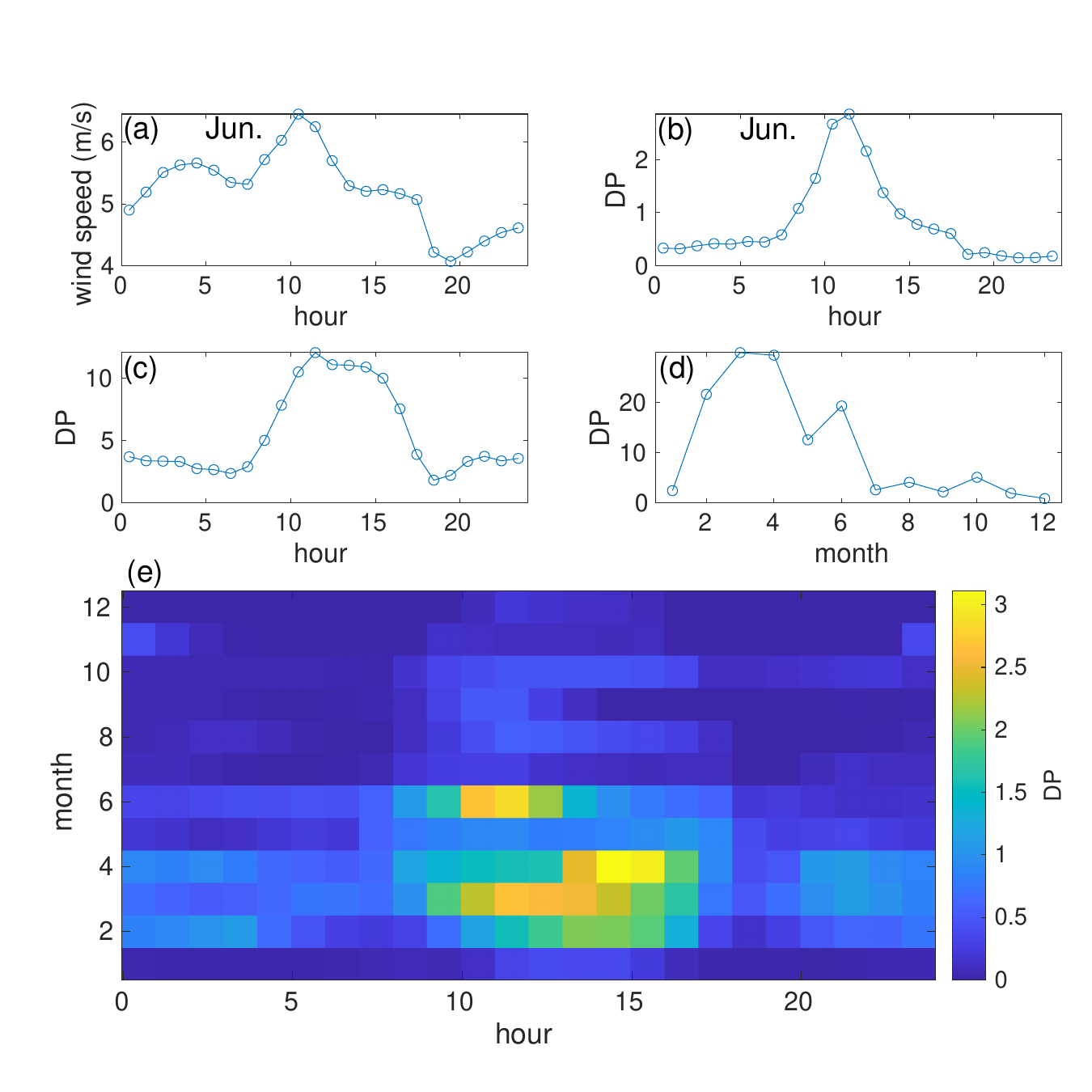}
  \end{center}
  \caption{{\bf Specific case.} The diurnal cycle of wind speed and DP at a specific location, 6.75\degree{E}, 30.75\degree{N}, between El Borma and Hassi Messaoud in Algeria; this location is indicated by the red ``$\times$'' symbol in Fig.~\ref{fig:dp_year}a. % Fig.~\ref{fig:dunes_map}.
  (a) The diurnal cycle of the wind speed (in m/s) during June 2021. (b) The diurnal cycle of DP (in vector units) during June 2021. Note the different curves of the wind speed and DP, due to the ability of DP to reflect mainly stronger winds. (c) The diurnal cycle of DP calculated based on the entire 2021 wind dataset. (d) The monthly DP during 2021. (e) DP as a function of the hour in the day and month of 2021. The results presented in panel (c) were obtained by summing the DP of the different months (i.e., summing each column of the matrix) while the results presented in panel (d) were obtained by summing over the hours (i.e., by summing over each column in the matrix).
    \label{fig:specific_location}}
\end{figure}

\begin{figure}
  \begin{center}
    \includegraphics[width=23pc]{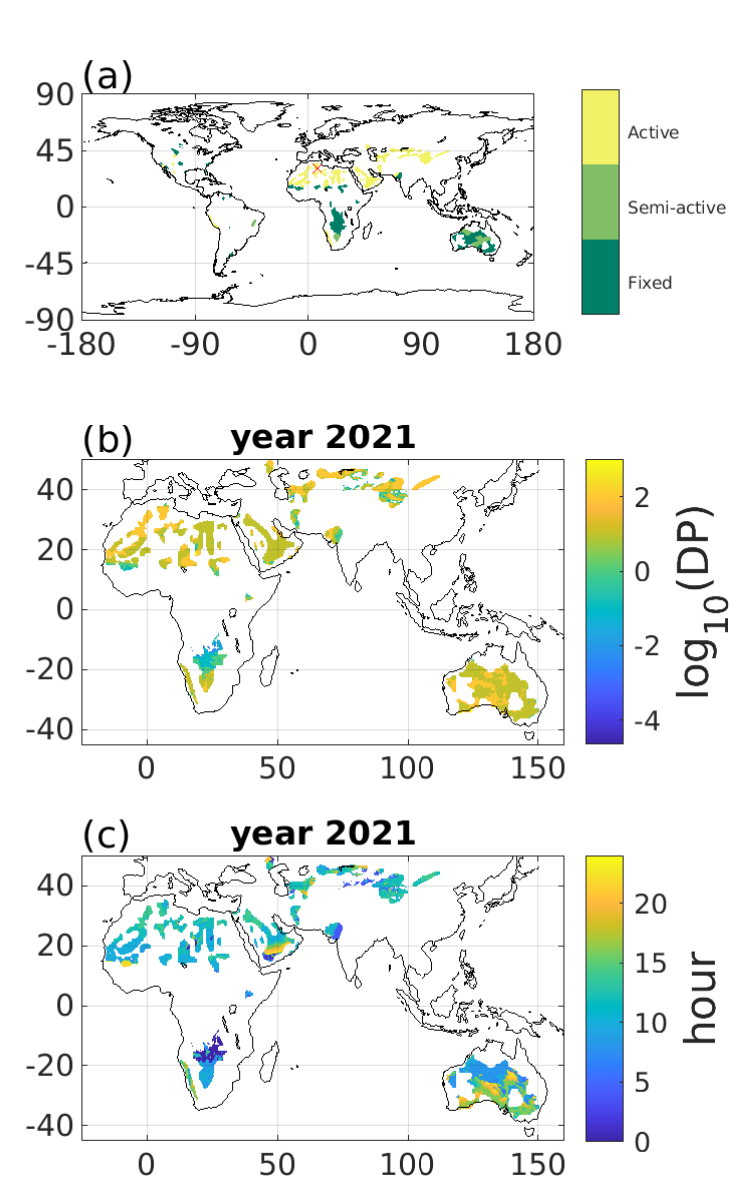}\\
  \end{center}
  \vskip -1cm
  \caption{{\bf Sand dune analysis.} (a) A map showing the sand dune regions over the globe classified into active (yellow), semi-active (light green), and stable (green) dunes. The red ``$\times$'' indicates the location of the data shown in Fig.~\ref{fig:specific_location}, after~\cite{Thomas-1997:sand}; see also~\cite{Ashkenazy-Yizhaq-Tsoar-2012:sand}. The dune activity can be defined according to mobility indexes \cite[e.g.,][]{Lancaster-1988:development,Tsoar-2005:sand,Thomas-Knight-Wiggs-2005:remobilization} or using a modeling approach \cite[e.g.,][]{Yizhaq-Ashkenazy-Tsoar-2007:why, Yizhaq-Ashkenazy-Tsoar-2009:sand}; see \cite{Abbasi-Opp-Groll-et-al-2019:assessment}. For example, according to the mobility index of \cite{Lancaster-1988:development} the mobility index $M$ is defined as the percentage of time the wind is above the threshold velocity over the ratio between the precipitation and evapotranspiration. A dune is defined as active when $M>200$, partially active when $50<M<200$, and inactive when $M<50$. (b) Annual (2021) $\log_{10}$DP over dune regions. (c) Time (hour) of maximum DP over dune regions during 2021.
    \label{fig:dp_year}}
\end{figure}

We first demonstrate our analysis (Fig.~\ref{fig:specific_location}) in a specific, representative location of 6.75\degree{E} and 30.75\degree{N}, in the sandy region of southeastern Algeria (the northern Sahara); this location is indicated by the red ``$\times$'' in Fig.~\ref{fig:dp_year}a. %\ref{fig:dunes_map}.
Fig.~\ref{fig:specific_location}a,b presents the diurnal cycle of the wind speed and DP during June 2021. It is clear that the wind is the strongest at the midday hours and that it is much weaker at night. The DP exhibits a more pronounced day-night contrast, as it enhances the effect of stronger winds; see Eq.~(\ref{eq:dp}) of the Methods section. Fig.~\ref{fig:specific_location}c shows the annual mean diurnal DP cycle, and also here, the stronger winds during the day are evident. We also present in Fig.~\ref{fig:specific_location}d the annual DP cycle, which indicates that the winds are stronger from February–June, during which the winds peak during the daytime (Fig.~\ref{fig:specific_location}e). Thus, we expect significant dune activity in these months, during the daytime.

Next, we present the DP over the main dune fields (Fig.~\ref{fig:dp_year}a); % (Fig.~\ref{fig:dunes_map});
 extended dune fields can be found mainly in Africa, Asia, and Australia, and thus, we first focus on these regions. The 2021 DP is presented in Fig.~\ref{fig:dp_year}b, and a comparison with Fig.~\ref{fig:dp_year}a % Fig.~\ref{fig:dunes_map}
suggests a link between the dunes' activity and DP, as was discussed in detail in previous studies~\cite[][]{Fryberger-1979:dune, Yizhaq-Xu-Ashkenazy-2020:effect}. Fig.~\ref{fig:dp_months} depicts the DP of the different months. Fig.~\ref{fig:dp_year}c presents the timing (the hour in the day) of the maximum DP in the year 2021; i.e., the time of the maximum in Fig.~\ref{fig:specific_location}c for different locations. The results of the different months are depicted in Fig.~\ref{fig:time_max_dp_dunes}. It is evident that over most dune areas, the maximum DP occurs during the daytime, and it is difficult to find locations in which the maximum DP occurs between midnight and early morning. We note that we did not find any clear relation between the dune activity (i.e., whether the dune is active or stable) and the timing (within the diurnal cycle) of maximum DP; the level of activity depends on the DP magnitude. The times of 2021’s maximum DP over the ocean and land are presented in Supplementary Figs.~\ref{fig:time_max_dp_annual}, \ref{fig:time_max_dp_ocean}, and \ref{fig:time_max_dp_land}. Over the ocean, there is no clear diurnal cycle (Fig.~\ref{fig:time_max_dp_ocean}), while over the land, the winds peak during the daytime, except in icy regions of Antarctica and Greenland and other high elevation regions. These findings will be discussed below. 

\begin{figure}
%\centerline{\includegraphics[width=\linewidth]{./Figures/time_of_max_dp_lat.pdf}}
\centerline{\includegraphics[width=\linewidth]{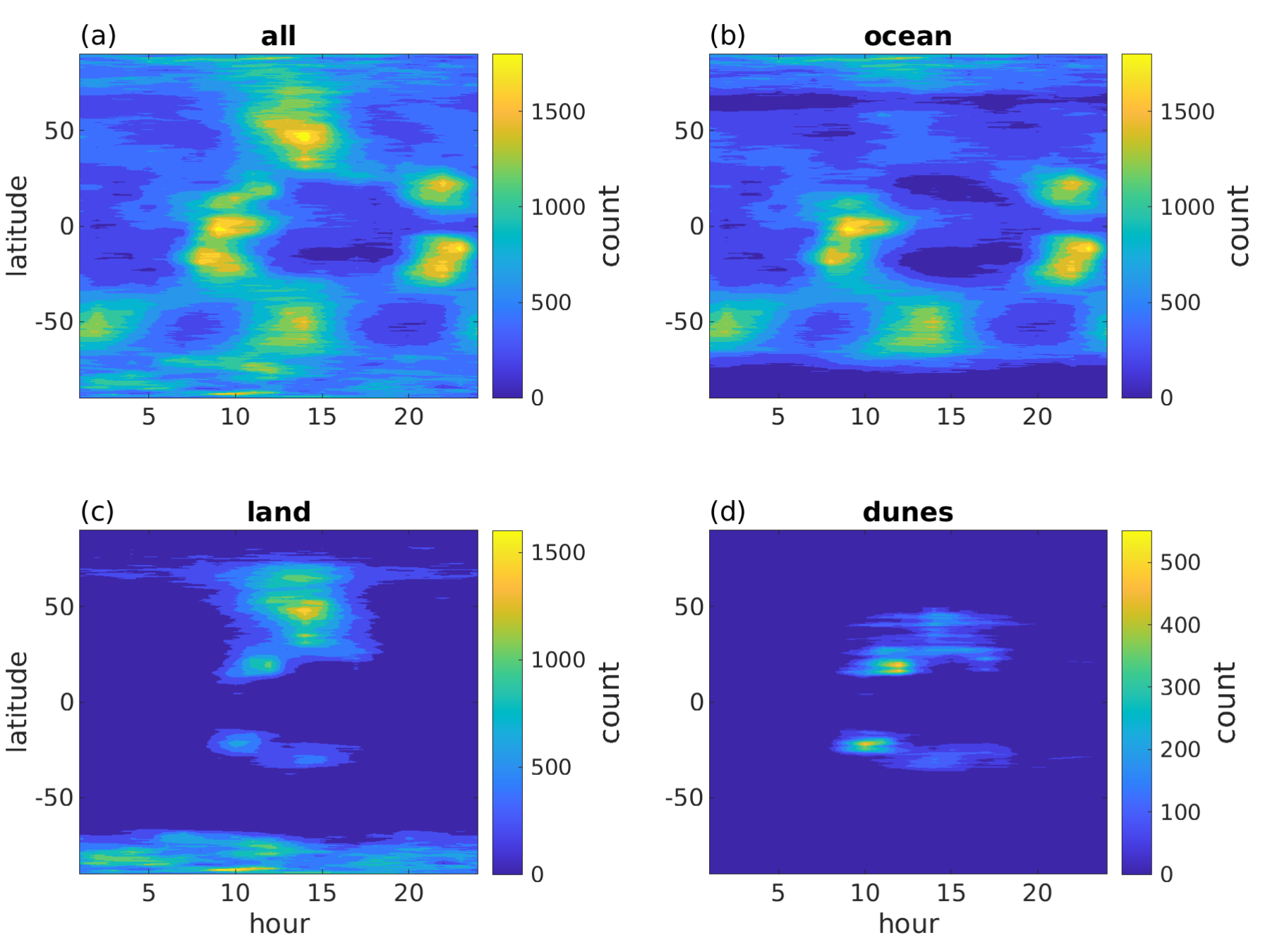}}
\caption{
  {\bf Histograms of the timing of maximum DP for 2021 versus latitude} for (a) the entire globe, (b) the ocean, (c) the land, and (d) the dune regions. Note the single peak during daytime over the land and over the dune regions and the double peak (whose timing varies with latitude) over the ocean. The results of the entire globe mainly reflect the results obtained for the ocean. }
\label{fig:time_of_max_dp_lat}
\end{figure}

Next, we concentrate on the statistics of the time (within a day’s 24 hours) at which the maximum DP occurs. More specifically, for each grid point and each month, we find the time (hour) of maximum DP. Then, we plot, for each latitude, the frequency of these hours for the entire latitude band, separately for the ocean, land, and dune regions within the latitude band (Fig.~\ref{fig:time_of_max_dp_lat}). The time of 2021’s maximum DP over land and dune regions is during the daytime, with maximum frequency around 14:00 over land and earlier (around 11:00) over the dune regions (Fig.~\ref{fig:time_of_max_dp_lat}c,d). An exception is Antarctica, over which there is no clear preferred timing of maximum DP within the diurnal cycle. To a lesser extent, this is also true in the high latitudes of the northern hemisphere ($\sim$60\degree{N}). This will be discussed below. As for the ocean (Fig.~\ref{fig:time_of_max_dp_lat}b), there is no clear maximum; at a latitude band of $\sim$50\degree{S}, there are two peaks, at midnight and at 14:00, while at another latitude band of $\sim$20\degree{S} to the equator,  we observe two peaks at $\sim$9:00 and $\sim$22:00. At other latitudes, there is no clear maximum. The results of the entire latitude band (Fig.~\ref{fig:time_of_max_dp_lat}a) mainly resemble, as expected, the results for the ocean, except for certain latitudes at which the relative ocean area is small (i.e., Antarctica).

\begin{figure}
  \includegraphics[width=36pc]{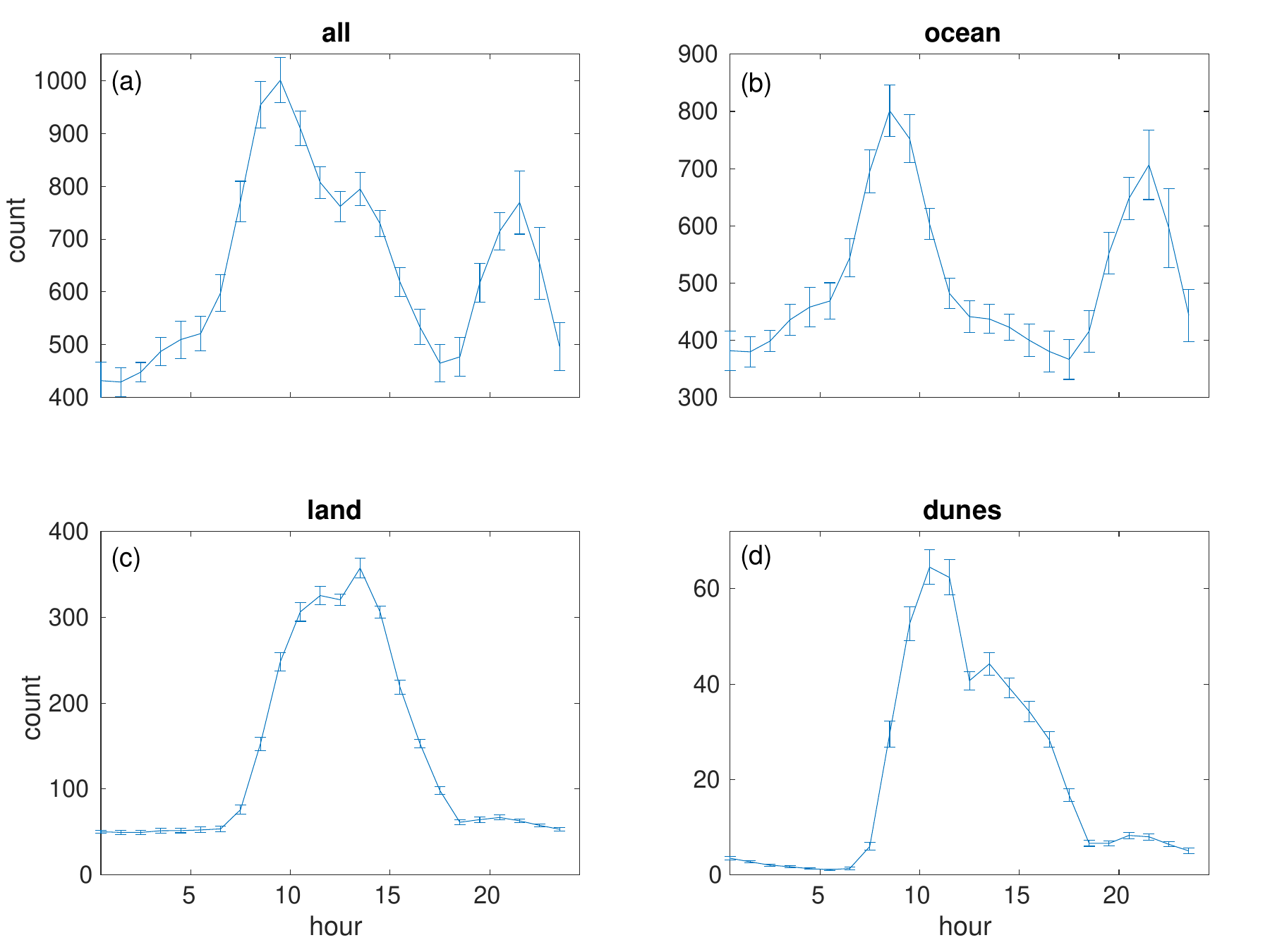}
  \caption{{\bf Histograms of the timing of maximum DP} averaged over 1950--2021 (mean $\pm$ std) for the (a) entire globe, (b) ocean, (c) land, and (d) dune regions. The mean and standard deviations of the pdfs were obtained by averaging over the pdfs of different years. The grid point area (shrinking poleward) was taken into account.
    \label{fig:time_of_max_dp_lat_years_mean}}
\end{figure}

\begin{figure}
%\centerline{\includegraphics[width=\linewidth]{./Figures/time_of_max_dp_lat_years.pdf}}
\centerline{\includegraphics[width=\linewidth]{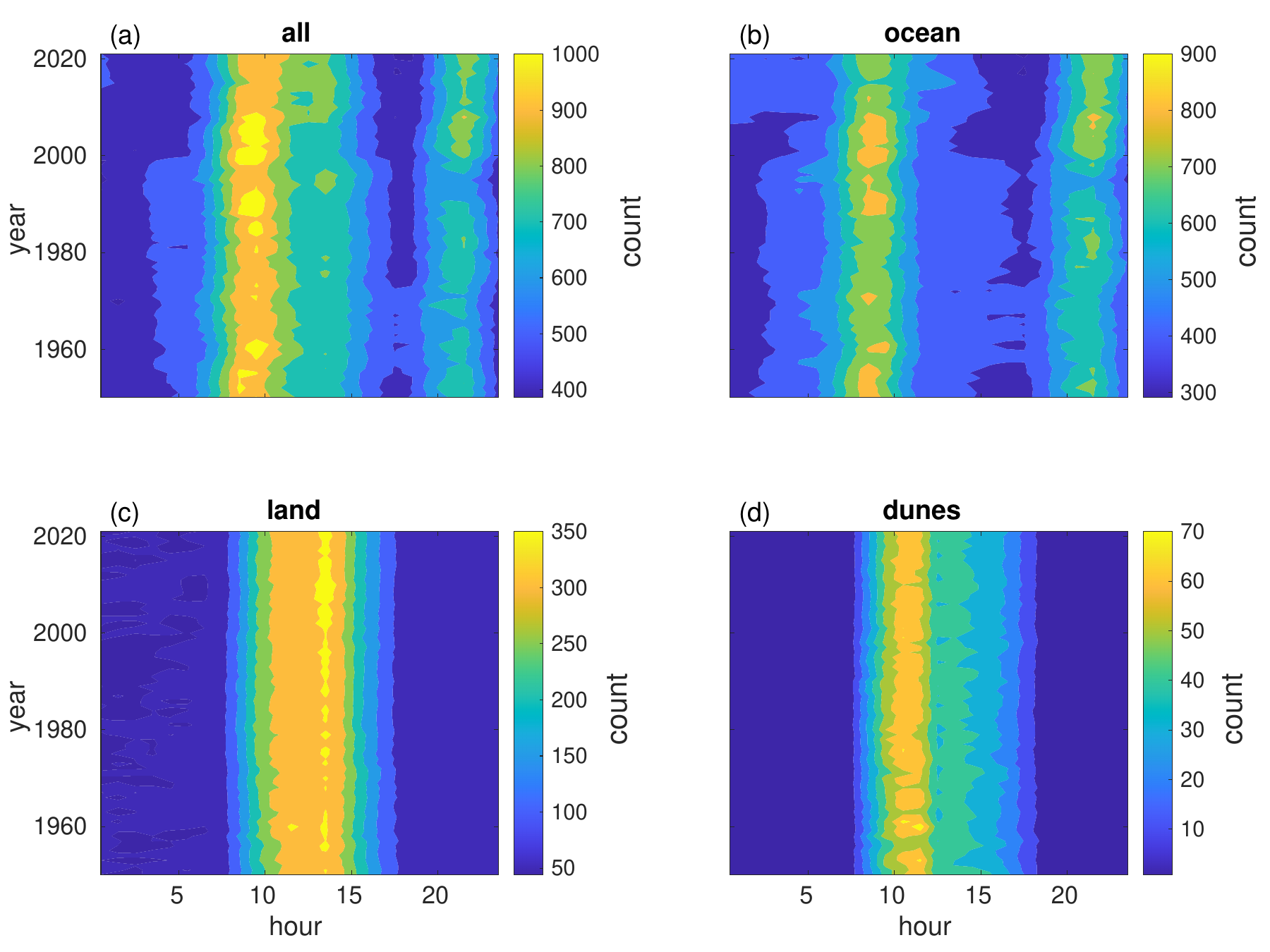}}
\caption{
  {\bf Histograms of the timing of maximum DP during 1950--2021} for the (a) entire globe, (b) ocean, (c) land, and (d) dune regions. Note that the histograms are relatively constant over time.  }
\label{fig:time_of_max_dp_lat_years}
\end{figure}

In the above, we concentrated on a specific year, 2021. In Fig.~\ref{fig:time_of_max_dp_lat_years_mean}, we summarize the results of 72 years, from 1950--2021. For every year, we constructed the meridional mean of Fig.~\ref{fig:time_of_max_dp_lat}, taking into account the effect of the decreasing grid area poleward and presented the temporal mean $\pm$ 1 std. In all the plots, the error bars are relatively small, indicating the stability of the results over time. For the dune regions (Fig.~\ref{fig:time_of_max_dp_lat_years_mean}d), it is clear that the DP peaks at $\sim$11:00 and that the surface winds are almost inactive during the nighttime hours, 18:00 to 7:00. The situation is similar for the land areas (Fig.~\ref{fig:time_of_max_dp_lat_years_mean}c), with the following differences: (i) the counts do not drop to zero during the nighttime hours when they are around 50. These nocturnal counts are due to the Antarctic timing of the maximum DP, which in this continent, occurs over the entire 24 hours. For the ocean (Fig.~\ref{fig:time_of_max_dp_lat_years_mean}b), in contrast to the single peak of the land and dune results, we observe two peaks, at 8:00 and 22:00. These two peaks reflect the low latitude results (see Fig.~\ref{fig:time_of_max_dp_lat}b), as this latitude range spans larger areas than the higher latitudes of the southern hemisphere ($\sim$50\degree{S}, Fig.~\ref{fig:time_of_max_dp_lat}b), for which there are two peaks, at midnight and at 14:00. The results of the entire globe mainly reflect, as expected, the results from the ocean. We also plotted the histograms of maximum DP times for all years from 1950-2021, and these indicate almost no variability with time (Fig.~\ref{fig:time_of_max_dp_lat_years}) for the land and dune regions and some minor variability for the ocean. 

\begin{figure}
%\centerline{\includegraphics[width=0.95\linewidth]{./Figures/time_of_max_dp_20min_data_1.pdf}}
\centerline{\includegraphics[width=0.95\linewidth]{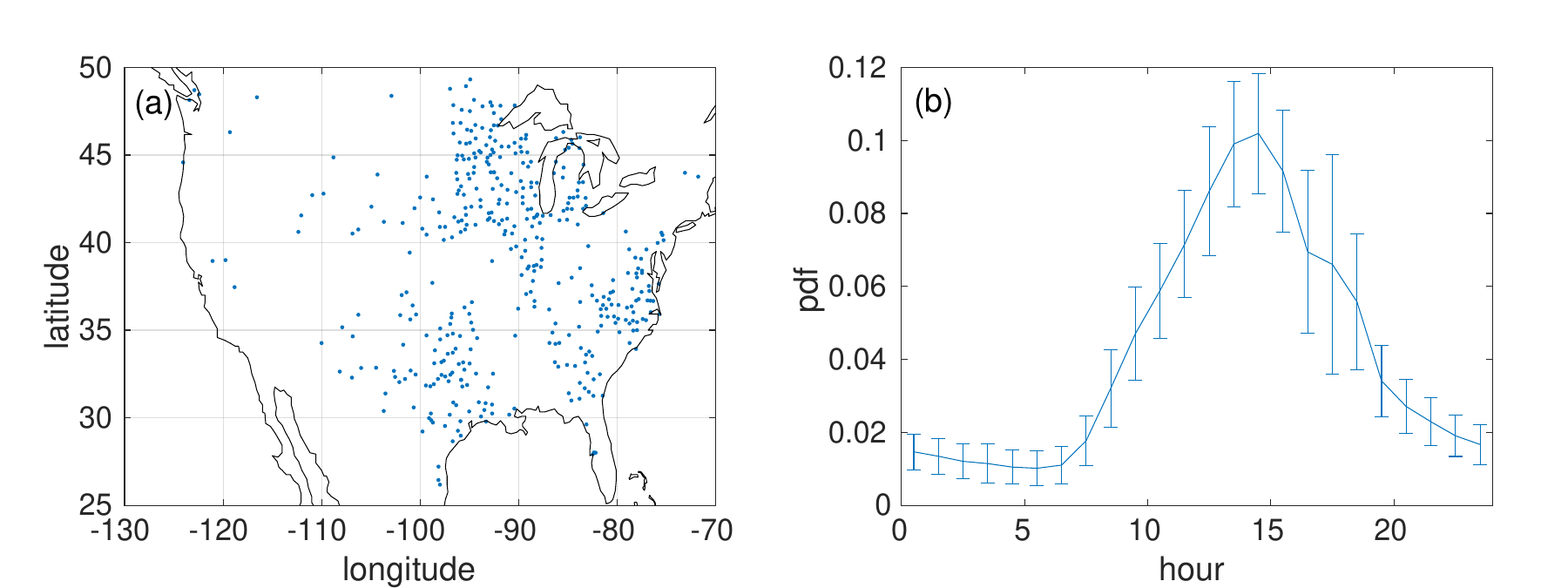}}
\caption{{\bf Meteorological (NCDC) measured wind data analysis.} (a) The locations of 406 USA stations. (b) Histograms of the timing (hour) of daily maximum wind speed. Most of the station data span the time period of 2009--2018.
  \label{fig:time_of_max_dp_20min_data}}
\end{figure}

The analysis we presented above is based on the ERA5 reanalysis data. To validate these results, we repeated the analysis based on wind data of 406 meteorological stations in the USA. The distribution of the stations is presented in Fig.~\ref{fig:time_of_max_dp_20min_data}a, and the mean $\pm$ std of the pdf of the time of daily maximum speed in the different stations is shown in Fig.~\ref{fig:time_of_max_dp_20min_data}b. Also here, as for the reanalysis data, the wind speed is much stronger during the daytime and peaks around 14:00. The relatively small standard deviations in the pdf indicate the results’ robustness. These results are consistent with Fig.~\ref{fig:time_max_dp_land}. 

Why are winds over land stronger during the day? This question was discussed in~ref.~~\cite[][]{Gunn-Wanker-Lancaster-et-al-2021:circadian}, in the context of a dune-field activity. These authors performed a field experiment at White Sands National Park (New Mexico, USA) and reported enhanced daytime sand and dust activity. They attributed this activity to stronger winds and drier conditions during the day. The atmospheric boundary layer extends to higher elevations after sunrise, as surface air warms and convects upward, making the air vertically unstable. They also found, based on 45 dune fields worldwide, a connection between surface winds and the diurnal temperature cycle.

Based on the paper discussed above, we suggest that the enhanced continental winds, during the daytime, are related to temperature-inversion, which, when present, disconnects the upper active atmospheric layers from the surface. Among the different types of inversions, we suggest that the ground inversion is the dominant one. The ground inversion is associated with the low heat capacity (low thermal inertia) of the ground, which cools rapidly after sunset. The low heat capacity is mainly due to the exponential decay of temperature with depth, such that only the upper few cms of the soil are affected by the daytime heating; soil’s smaller heat capacity (by about a factor of two), relative to that of seawater, plays a less significant role. The lower ground temperature at night cools the air above it, usually resulting in cold (and dense) air underlying warmer (and lighter) air, leading to stable atmospheric conditions, isolating the surface from a possibly active upper layer. After sunrise, the ground warms up within a few hours, warming the air above it and resulting in vertical convection, eventually connecting the surface with the upper, more active atmospheric levels.
% Over desert regions where air subsidence occurs, subsidence inversion may also play a role in disconnecting the upper atmospheric levels from the lower levels, leading to weaker surface winds.
Over the ocean, the sea surface temperature hardly varies over the time scale of one day, due to the larger specific heat of seawater and, more importantly, due to the mixed layers of depths of tens of meters. For this reason, we do not observe a clear timing for the maximum wind over the ocean; we do not have an explanation for the two peaks observed over oceanic regions (Figs.~\ref{fig:time_of_max_dp_lat}, \ref{fig:time_of_max_dp_lat_years_mean}).

The air stability close to the surface may be examined through the sensible heat flux, which is proportional to the temperature difference between the air and the surface (water, soil, ice, etc.) and to the surface wind speed. The sensible heat flux is the conductive heat flux emitted from the earth’s surface to the atmosphere and plays an important role in the energy budget at the earth’s surface. Given the above, over land, during the daytime, the sensible heat flux is negative (i.e., the temperature of the air at the surface is warmer and possibly lighter than the air above it, possibly leading to unstable conditions and to the connection of the surface with the upper active atmospheric levels), whereas at night, it is positive (i.e., the temperature of the air at the surface is colder and heavier than the air above it, leading to stable conditions and to the disconnection of the surface from the upper active atmospheric levels). The surface’s small heat capacity leads to a large difference between the maximum sensible heat flux at night and the minimum flux during the daytime.

\begin{figure}
%\centerline{\includegraphics[width=0.95\linewidth]{./Figures/SH-maps_1.pdf}}
\centerline{\includegraphics[width=0.95\linewidth]{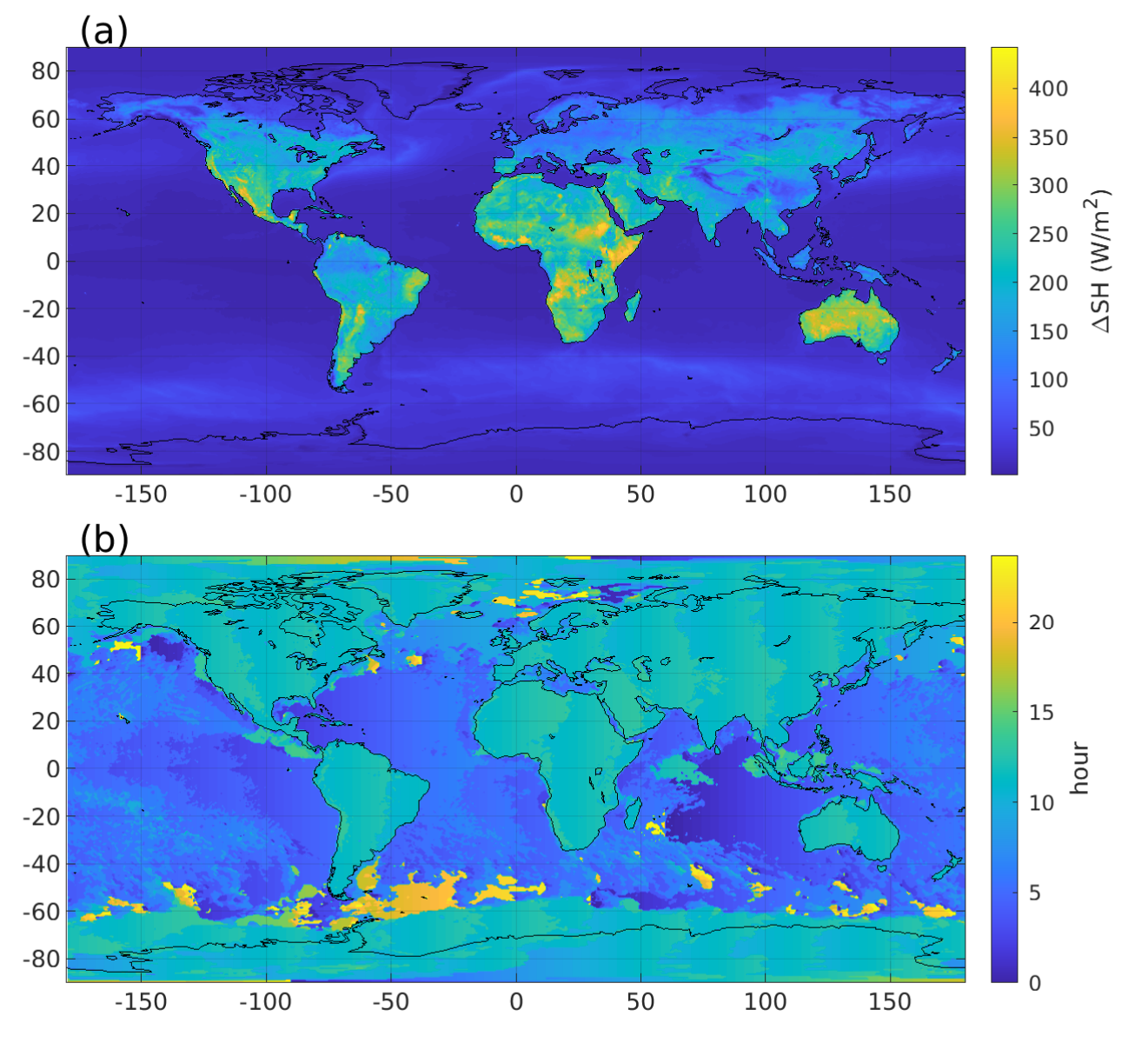}}
\caption{{\bf Sensible heat flux analysis.} (a) Map showing the difference between the daily maximum and daily minimum ERA5 sensible heat flux in 2021. Note the much larger values over land in comparison to the ocean, indicating a maximum in wind activity during the daytime. Exceptions are Antarctica and Greenland. (b) The hour (within the day) of minimum sensible heat flux in 2021.
  \label{fig:SH_maps}}
\end{figure}

Over the ocean, this difference is expected to be small, due to the small variations in sea-surface temperature during the day, associated with the mixed layer’s large heat capacity. In Fig.~\ref{fig:SH_maps}a, we present the difference between the maximum and minimum sensible heat flux during the daytime, $\Delta {\rm SH}$, and it is evident that indeed the difference is much larger over the land than over the ocean; this supports our explanation regarding the stability of the air over land and over the ocean and during the day and night hours. This is true, despite the stronger winds over the ocean that should have increased the sensible heat flux.

Interestingly, the night-day difference in sensible heat flux over Antarctica and Greenland is similar to that of the ocean and may explain the observation that maximum wind occurs over the entire 24 hours of the day (Fig.~\ref{fig:time_max_dp_annual}, \ref{fig:time_max_dp_land}, \ref{fig:time_of_max_dp_lat}). This may be related to the high altitude of Antarctica and Greenland and to the high albedo of the ice/snow, which reduces the absorbed solar radiation; see more discussions in the next paragraph. In Fig.~\ref{fig:SH_maps}b, we present the timing of the minimum sensible heat flux, which can be related to greater vertical instability and, thus, stronger winds. Indeed, over land, the minimum occurs during the daytime, while over the ocean, the minimum occurs during the last hour of the night. 

There is no clear timing for the maximum wind over Antarctica and Greenland, and to a lesser extent, the daytime maximum of surface winds is less clear at the high elevation continental latitudes. This might be related to the extended time of either the day or night at the high latitudes or to the high surface elevations of Antarctica and Greenland, which makes the surface too difficult to disconnect from the high--altitude atmospheric activity. The higher albedo of the ice results in less absorbed shortwave radiation, which can weaken the effect of surface heating/cooling that influences the daytime maximum wind over the land. The results regarding the sensible heat flux over these regions (Fig.~\ref{fig:SH_maps}b) are consistent with the results regarding the timing of the maximum wind.

\subsection{The temporal trends of wind.}\label{sec3_2}

To complement the analysis of the timing of the maximum wind speed, we present below the results regarding the temporal variability of DP, wind power, and wind speed. Previous studies indicated a significant increase in surface wind speed over the ocean with time~\cite[][]{Young-Zieger-Babanin-2011:global,Young-Ribal-2019:multiplatform,Yu-Zhong-Sun-2020:climatology} and, to a lesser degree, over land~\cite[][]{Yu-Zhong-Sun-2020:climatology}. In addition, a general increase in wind speed was reported based on different reanalysis models~\cite[][]{Wohland-Omrani-Witthaut-et-al-2019:inconsistent} although, in some locations, the trends were not consistent; here, we focus on trends based on the relatively new and fine resolution reanalysis, the ERA5, and present results of the DP, which, to our knowledge, have not been previously presented. We note that reanalysis data does not always accurately reflect reality. More specifically, the wind speed of the ECMWF reanalysis underestimates real winds, while that of the NCAR/NCEP reanalysis overestimates real winds~\cite[][]{Ashkenazy-Yizhaq-Tsoar-2012:sand, Stopa-Cheung-2014:intercomparison, Yizhaq-Xu-Ashkenazy-2020:effect}. Below, we further analyze the difference between the ERA5 and measured data over the USA. We also show that although the DP magnitude is not well captured by the ERA5, the DP trends probably reflect real trends.

\begin{figure}
%\centerline{\includegraphics[width=0.75\linewidth]{./Figures/time_trends_dp.pdf}}
\centerline{\includegraphics[width=0.75\linewidth]{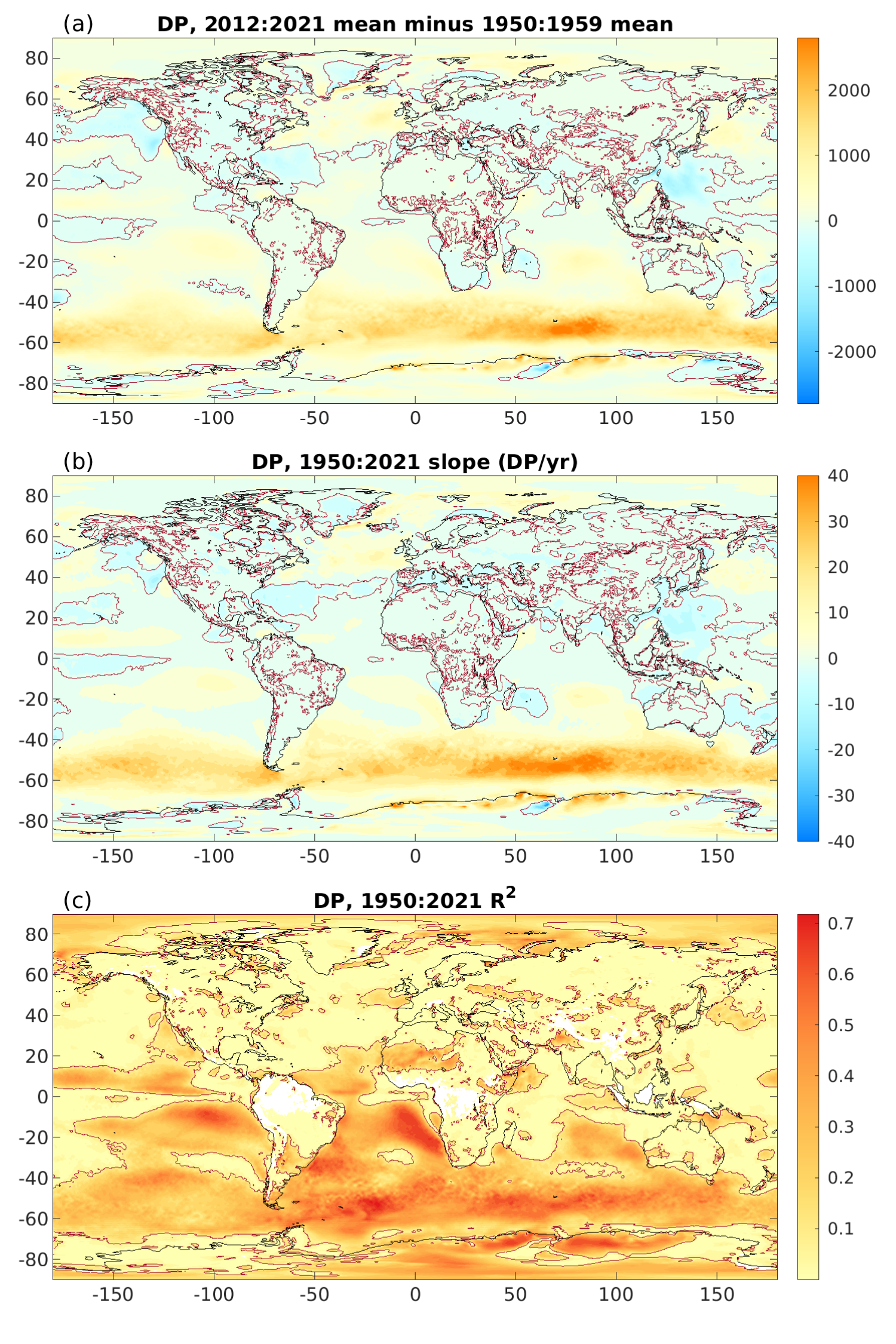}}
\caption{{\bf Temporal trends in ERA5 DP.} (a) The difference between the mean DP of the last ten years (2012--2021) and the first ten years (1950--1959). The red contour line indicates the zero level, while the black contour line indicates the sea-land boundaries. Note the high increase over the ocean. (b) The linear slope of the annual mean DP over 1950--2021. The red contour line indicates the zero level. (c) The $R^2$ of the slopes that are presented in panel b. The red contour line indicates the 99\% confidence level of the slope (which is approximately $R^2\approx0.09$), above which the slope can be regarded as significant.
  \label{fig:time_trends_dp}}
\end{figure}

To estimate whether and by how much the wind varied over the years, we subtract (at each grid point) the mean DP of the first 10 years of the ERA5 (1950--1959) from the mean of the last 10 years (2012--2021). The results are shown in Fig.~\ref{fig:time_trends_dp}a and indicate that most of the globe experienced an increase in DP, especially over the ocean. The increase is more pronounced over the ocean than over continental regions. In addition, we estimated, at each grid point, the rate of the increase/decrease in DP over time, i.e., we have performed a linear regression for each time series (of each grid point) and considered the corresponding slope as the rate of the local grid point. The results are shown in Fig.~\ref{fig:time_trends_dp}b where also here the rate (slope) is positive over most of the world’s area, with more extensive regions with a positive slope and a significantly larger increase rate over the ocean. In some oceanic regions (e.g., the Southern Ocean), the increase exceeds 30 vector units of DP per year. This increase is consistent with previous studies~\cite[][]{Hande-Siems-Manton-2012:observed, Young-Ribal-2019:multiplatform}.

Interestingly, extended oceanic regions experienced a decline in DP, including the western tropical Pacific, the central equatorial Pacific, the sub-tropical North Atlantic Ocean, and the Mediterranean Sea. To estimate the significance of the increase, we have also calculated the $R^2$ of the slope (Fig.~\ref{fig:time_trends_dp}c). To further estimate at what value of $R^2$ one can consider the slope as significant, we have randomly selected $10^5$ grid points and shuffled the time series of each of these points. The mean slope of these shuffled time series is around zero. Then, we calculated the $R^2$ of these shuffled time series and found the 99\% percentile of these $R^2$ values; this percentile is $\sim0.09$, meaning that a time series with a $R^2>0.09$ has only a 1\% probability to be associated with a random time series. It is apparent from Fig.~\ref{fig:time_trends_dp}c that extended oceanic regions experienced a significant increase in DP, while most of the continental regions have $R^2<0.09$. Continental regions with a significant increase in DP include parts of the Sahara/Sahel and regions in South America. We have performed similar analyses on wind speed (Fig.~\ref{fig:time_trends_spd}) and on the cube of the wind speed (which is proportional to the wind power) and found similar results.

\begin{figure}
%\centerline{\includegraphics[width=0.95\linewidth]{./Figures/dp_time.pdf}}
\centerline{\includegraphics[width=0.95\linewidth]{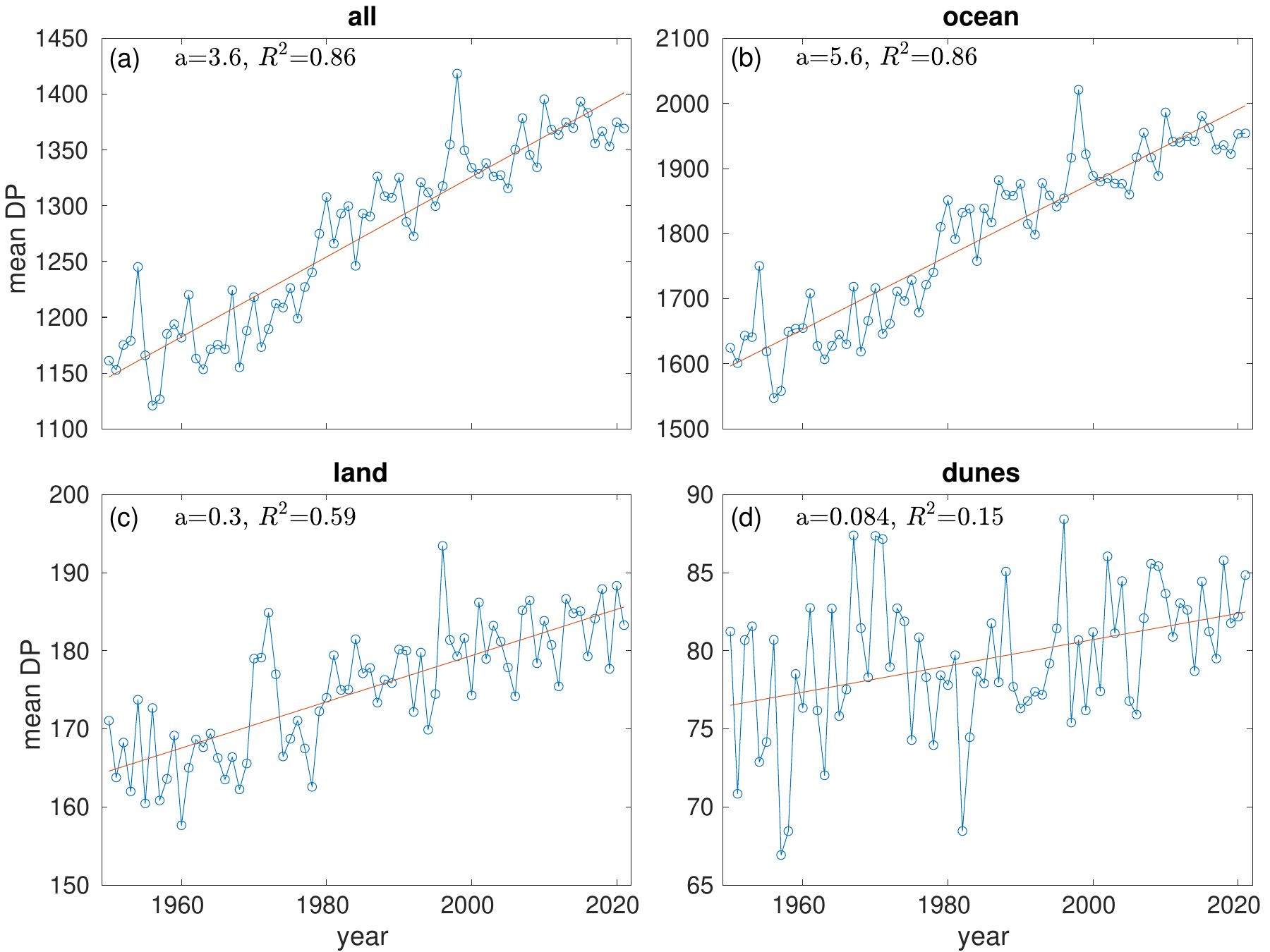}}
\caption{{\bf Temporal trends in ERA5 DP (1950--2021).} Averaged over (a) the entire globe, (b) the ocean, (c) the land, and (d) sand dune regions. The red line indicates the fitted linear trend and the corresponding slope $a$ and $R^2$. Note the significant increase over the ocean (b) and a much weaker increase over the sand dune regions.
  \label{fig:dp_time}}
\end{figure}

We next present the annual DP versus time (1950--2021) averaged over: (i) the entire globe, (ii) the ocean, (iii) the land, and (iv) dune regions; see Fig.~\ref{fig:dp_time}. While over all regions the increase is significant, the increase over the ocean is the largest and most significant, the increase over land is less significant and less pronounced, and the increase over the dune regions is the least significant and pronounced. Indeed, the increase in DP over most of the continental regions is not significant, and in some regions like Greenland, North America, and China, DP decreases with time; see Fig.~\ref{fig:time_trends_dp}. Such a decrease in DP is consistent with studies regarding sand dunes in China~\cite[][]{Mason-Swinehart-Lu-et-al-2008:limited,Li-Liu-Su-et-al-2015:changes,Wang-Dong-Yan-et-al-2005:wind,Zhang-Fan-Cao-et-al-2018:response}, in the Canadian prairies~\cite[][]{Hugenholtz-Wolfe-2005:recent,Hugenholtz-Bender-Wolfe-2010:declining}, and in the world's largest sand island in Eastern Australia~\cite[][]{Levin-2011:climate,Yizhaq-Ashkenazy-Levin-Tsoar-2013:spatiotemporal}. We have performed a similar analysis for the wind speed (Fig.~\ref{fig:spd_time}) and found similar results, although here, the significance of the increase over the land is similar to that over the dunes. 

\begin{figure}
%\centerline{\includegraphics[width=0.95\linewidth]{./Figures/era5_ncdc_us_eu.pdf}}
\centerline{\includegraphics[width=0.95\linewidth]{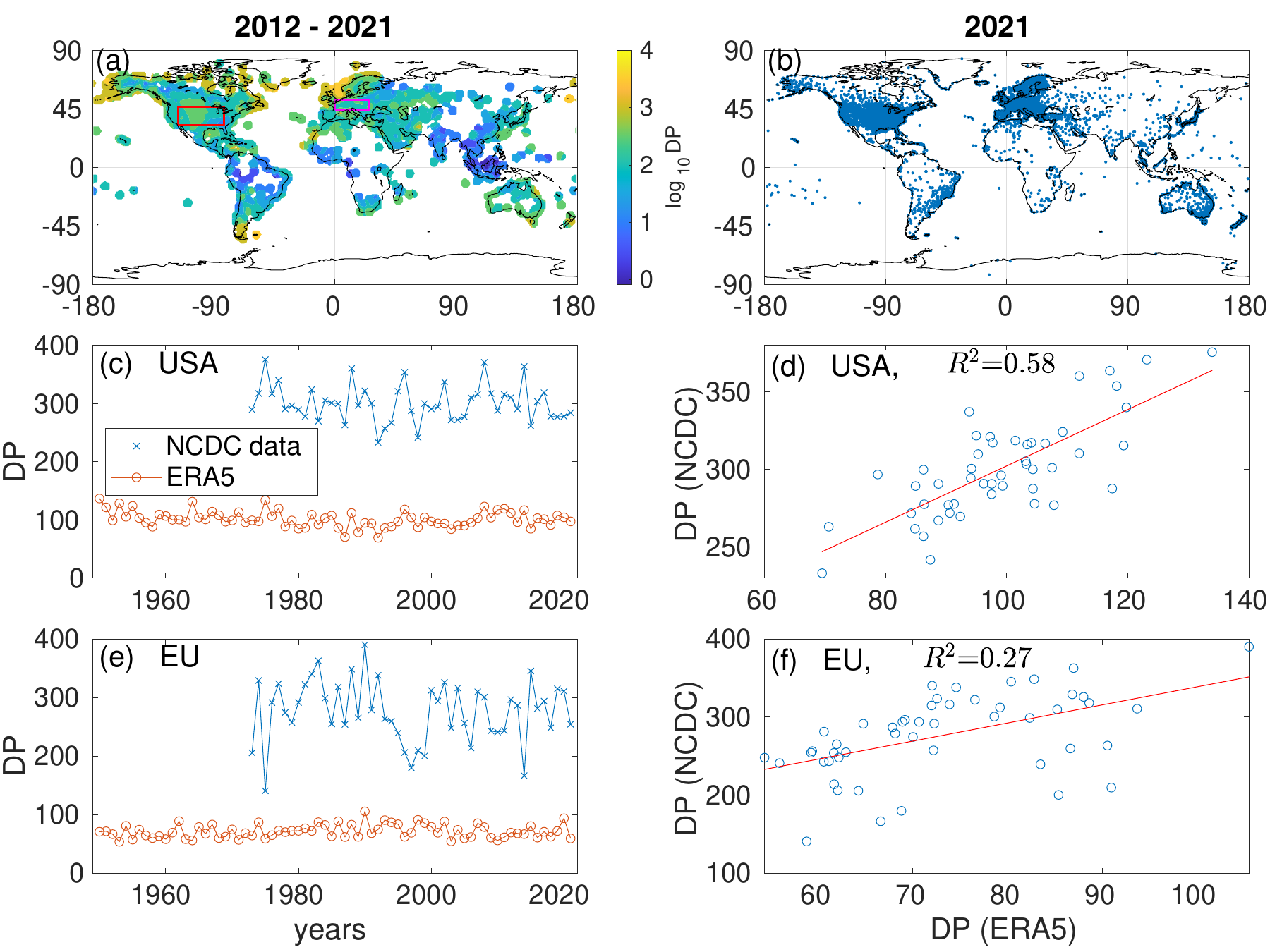}}
\caption{{\bf Meteorological NCDC (measured) DP analysis.} (a) A map showing the $\log_{10}\rm{DP}$ of the mean DP over the past 10 years (2012--2021). (b) Locations of 6718 stations that were used in the analysis of 2021. The mean number of stations during the period 2012--2021 is 6127$\pm$929. (c) The mean DP within the USA (indicated by the red rectangle in panel a) is based on measured NCDC data (blue) and on the ERA5 (red). The ERA5 DP is smaller, approximately, by a factor of 3 than that of the measured data. Still, the pattern of the two curves is similar. (d) NCDC DP versus the ERA5 DP for 1973--2021; the correlation coefficient, $R^2=0.58$, indicates non-negligible correlations. (e) Same as panel c for a region within Europe (indicated by the magenta rectangle in panel a). (f) Same as d for a region within Europe. Note the smaller correlation coefficient, $R^2=0.27$. Also here the ERA5 DP is smaller by about a factor of 3 than that of the measured data.
  \label{fig:era5_ncdc_us}}
\end{figure}

We have used the NCDC meteorological data to validate the trends observed in the ERA5; see Fig.~\ref{fig:era5_ncdc_us}. A map showing the mean DP over the past 10 years (2012--2021) is depicted in Fig.~\ref{fig:era5_ncdc_us}a. We focus on regions within the USA and Europe (indicated by the red and magenta rectangles in Fig.~\ref{fig:era5_ncdc_us}a), as there are many stations within these regions (Fig.~\ref{fig:era5_ncdc_us}b). Generally speaking, the DP of the ECMWF reanalysis (like the ERA5) underestimates the one that is based on measured data~\cite[][]{Ashkenazy-Yizhaq-Tsoar-2012:sand}. Fig.~\ref{fig:era5_ncdc_us}c,e shows the mean DP over the USA and Europe based on the ERA5 and NCDC data. First, there is no clear trend in DP in either data source. Second, the DP of the ERA5 is smaller than the DP of the NCDC by about a factor of 3. Yet, the pattern of the time series of the two data sources is similar (Fig.~\ref{fig:era5_ncdc_us}d,f), indicating that the trends of the ERA5 are most likely reliable, but their actual value is not. We also perform a similar analysis for the wind speed (Fig.~\ref{fig:era5_ncdc_us_spd}), but here, there is a decreasing trend in the wind speed using the NCDC and a much weaker trend in the ERA5. The match between the two sources is good in the last 15 years (or so) of the data (Fig.~\ref{fig:era5_ncdc_us_spd}a,b) and the difference between the NCDC data and the ERA5 is much smaller than the difference in the DP time series (Fig.~\ref{fig:era5_ncdc_us}c,e). This indicates that the ERA5 does not accurately capture the extreme winds, as DP signifies strong winds.
%We note that the higher number of stations in this region (Fig.~\ref{fig:era5_ncdc_us_spd}b) may affect the results of the NCDC data although the relatively high correlation between the ERA5 and the NCDC (Fig.~\ref{fig:era5_ncdc_us}d) suggests that this effect is small. 

The increase in DP and wind speed over time may be explained in different ways that were previously discussed at length~\cite[][]{Wohland-Omrani-Witthaut-et-al-2019:inconsistent}. First, the observed trends in the reanalysis data may be due to biases and measurement techniques that varied over time~\cite[][]{Cardone-Greenwood-Cane-1990:trends, Ward-1992:provisionally, Ward-Hoskins-1996:near, Thomas-Kent-Swail-et-al-2008:trends}, including the increase in heights of anemometers, the use of measured wind speed instead of estimated wind speed, and higher sampling rate as time advanced~\cite[][]{Wohland-Omrani-Witthaut-et-al-2019:inconsistent}. Yet, the increase of wind speed over the ocean was reported based on measured data~\cite[][]{Young-Zieger-Babanin-2011:global,Young-Ribal-2019:multiplatform,Yu-Zhong-Sun-2020:climatology}, and indeed, such an increase was found here. The relatively high correlation between the DP of the NCDC and the ERA5 (Fig.~\ref{fig:era5_ncdc_us}d and to a lesser degree Fig.~\ref{fig:era5_ncdc_us}f) also indicates that the trends in the ERA5 are not artifacts. Thus, it is possible that the new advanced reanalysis ERA5, which we analyzed here, more accurately reflects reality and that the increasing trends in wind speed are indeed real. Previous studies had reported the weakening of surface terrestrial winds and attributed it to an increased roughness due to urbanization and changes in vegetation~\cite[][]{Vautard-Cattiaux-Yiou-et-al-2010:northern, Wu-Zha-Zhao-et-al-2018:changes}. However, a more recent study reported a decreasing trend in terrestrial wind activity till 2010 followed by a more rapid increasing trend~\cite[][]{Zeng-Ziegler-Searchinger-et-al-2019:reversal} and suggested that the observed trends are associated with large-scale ocean-atmosphere decadal variability.

We also analyzed the trends of the sensible heat flux of the ERA5 (Fig.~\ref{fig:SH_time}). Here, the trends are not similar to the trends in DP and wind speed. Yet, there is an increasing trend over the ocean. In addition, there is a difference between the trends in the first phase of the data (1950--1980) and the last part of the data (1980--2021); this may be related to the two phases of the ERA5, namely, 1950--1978 and 1979 to the present day.  

\section{Summary and Discussion.}\label{sec4}

In this study, we investigated two aspects of the wind field---the diurnal cycle of the wind and trends in wind activity over the past 70 years or so. We considered three measures of the wind: (i) wind drift potential (DP), which captures extreme wind activity and is used as a measure for potential sand transport over sand dunes, (ii) wind speed, and (iii) wind power, which is proportional to the cube of the wind speed. We mainly focus on the timing of the DP maximum and show that over land, especially over the dune regions, the wind peaks during the daytime, whereas it significantly weakens at night; these results are consistent with a recent study regarding the sand activity in the area of White Sands National Park in the US~\cite[][]{Gunn-Wanker-Lancaster-et-al-2021:circadian} and with studies regarding sand and dust activity in the Taklimakan and Gurbantunggut Deserts~\cite[][]{Yang-He-Mamtimin-et-al-2013:diurnal, Yang-He-Liu-et-al-2018:saltation,  Zhou-Mamtimin-Pan-et-al-2019:relationship}. We attributed the enhanced winds, over land and during the daytime, to the small heat capacity of the ground, which warms and cools quickly, resulting in convection during the daytime and inversion at night; this leads to active upper atmospheric levels that are disconnected from the surface at night, which eventually leads to weaker nighttime surface winds. Over the ocean, there is no clear diurnal cycle, most likely due to the almost constant sea surface temperature on time scales of 24 hours. These results are supported by the difference between the daily maximum and the daily minimum of sensible heat flux.
%We observed two times over the 24 hours during which the wind is maximal, occurring at different hours at different latitude bands.
There is no clear diurnal cycle over icy regions such as Antarctica and Greenland, which is consistent with the small sensible heat flux difference over these regions. We did not observe significant temporal variations in the timing of the maximum wind from 1950 to 2021; this is in contrast to the increasing temporal trends in DP, wind speed, and wind power. 

We also analyzed the trends in DP, wind speed, and wind power during the period 1950--2021; we analyzed these trends over the entire globe, the ocean, the land, and sand dune regions. Increasing trends in wind speed were reported in several studies, both based on real data~\cite[e.g.,][]{Young-Zieger-Babanin-2011:global,Young-Ribal-2019:multiplatform,Yu-Zhong-Sun-2020:climatology} and reanalysis data~\cite[e.g.,][]{Wohland-Omrani-Witthaut-et-al-2019:inconsistent, Yu-Zhong-Sun-2020:climatology}; these previous studies did not perform the analysis on the ERA5 data using DP and did not focus on dune regions. We find significantly increasing wind activity (as reflected by DP, wind speed, and wind power) over the ocean, especially over the Southern Ocean, less significant increasing trends over land, and even less significant increasing trends over the dune regions. We note, however, that in some specific locations, the wind has weakened over the past decades. The overall increasing trend may be the result of biases and measurement techniques that have changed over time. Yet, previous studies reported increasing wind speed over the ocean~\cite[][]{Young-Zieger-Babanin-2011:global,Young-Ribal-2019:multiplatform}, especially over the Southern Ocean~\cite[][]{Yu-Zhong-Sun-2020:climatology}, strengthening the validity of our results. Results that are based on measured NCDC data indicate that the ERA5 underestimates real values, while the temporal trends are more reliable. Our results show that the increasing trend in DP over the dunes is weak, most likely indicating no (or weak) significant increase in dune mobility in the past 70 years or so; still, the above is valid on average, such that particular dune regions might experience a decreasing trend in wind activity. Winds are associated with pressure gradients, which are closely related to temperature gradients, which are, in turn, related to ocean-atmosphere interactions/oscillations~\cite[][]{Zeng-Ziegler-Searchinger-et-al-2019:reversal}; since the surface temperature has increased over the past decades, and in a spatially uneven manner, the temperature gradients may also have varied, leading to changes in wind activity.

In a recent study, \cite{Gunn-East-Jerolmack-2022:century} analyzed one of the Climate Model Intercomparison Project (CMIP6) models using different future Shared Socioeconomic Pathway (SSP) scenarios and predicted reduced activity and stagnation during the 21 century of the currently active dunes. According to this study, the stagnation process started around 1970. These findings are not fully consistent with the results we reported here, that the DP exhibited a slight increase over the dune regions from 1950 to 2021. The difference might be due to the different nature of ERA5 and the CMIP6 models, where the former is adjusted to the observed data every several hours while the latter starts from 1850 initial conditions and then integrated forward in time. 

% Interestingly, it was recently discovered that in the Gale Crater on Mars, the winds are stronger at night~\cite[][]{Baker-Newman-Sullivan-et-al-2021:diurnal}, possibly due to some local effects. There are also some locations on earth in which the winds are stronger at night, so it may be that in other places on Mars outside the Gale Crater, the winds are stronger during the day. Following the above, we note that in other bodies in the solar system (such as Titan, the moon of Saturn), the wind’s diurnal cycle may be different from earth due to the different physical conditions.

Winds play a central role in the climate system, on both global and local scales. Thus, it is important to study the wind’s characteristics. Moreover, winds are the driving force of dunes and determine the dune’s activity level and morphology \cite[][]{Sun-Gao-2022:geomorphology}, and dust emission. Our results indicate, as was proposed in a previous study~\cite[][]{Gunn-Wanker-Lancaster-et-al-2021:circadian}, that dunes are active during the daytime and are almost completely inactive at night. Our results may have important consequences for wind energy production over the land and the ocean.

\section*{Acknowledgments}





\bibliographystyle{elsarticle-harv}
%\bibliography{all}




% \end{thebibliography}

\section*{Figures}

\appendix
\clearpage\newpage
\section*{Supplementary Figures}
\clearpage\newpage

%%% Supplementary figures
\setcounter{figure}{0}
\renewcommand{\thefigure}{S\arabic{figure}}

\begin{figure}
%\centerline{\includegraphics[width=\linewidth]{./Figures/dp_months1.pdf}}
\centerline{\includegraphics[width=\linewidth]{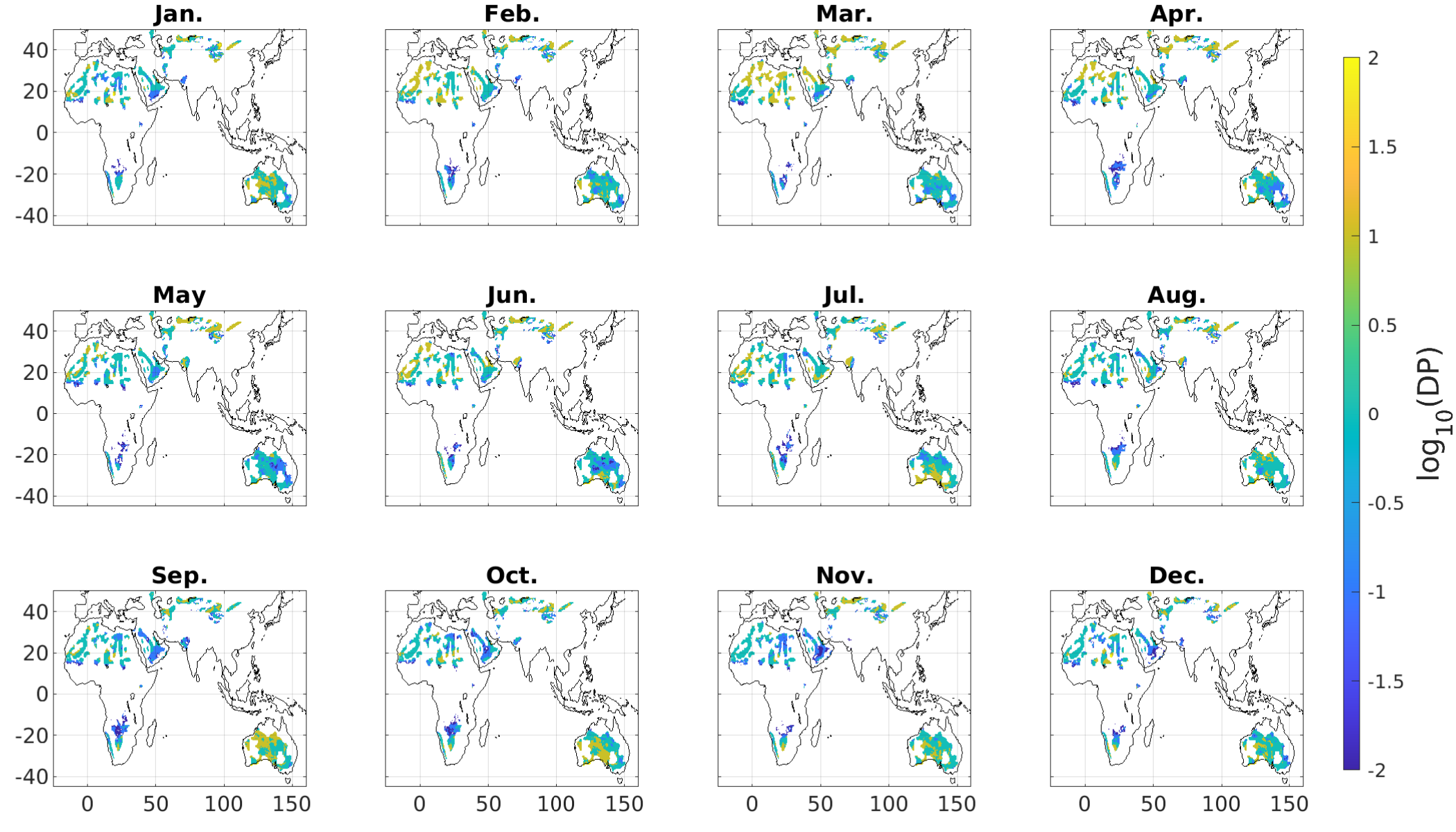}}
\caption{
  {\bf Monthly $\log_{10}$DP over dune regions during 2021.} The annual DP is the sum of all months. The results are based on the ERA5.
\label{fig:dp_months}}
\end{figure}

\begin{figure}
%\centerline{\includegraphics[width=\linewidth]{./Figures/time_of_max_dp_dunes.pdf}}
\centerline{\includegraphics[width=\linewidth]{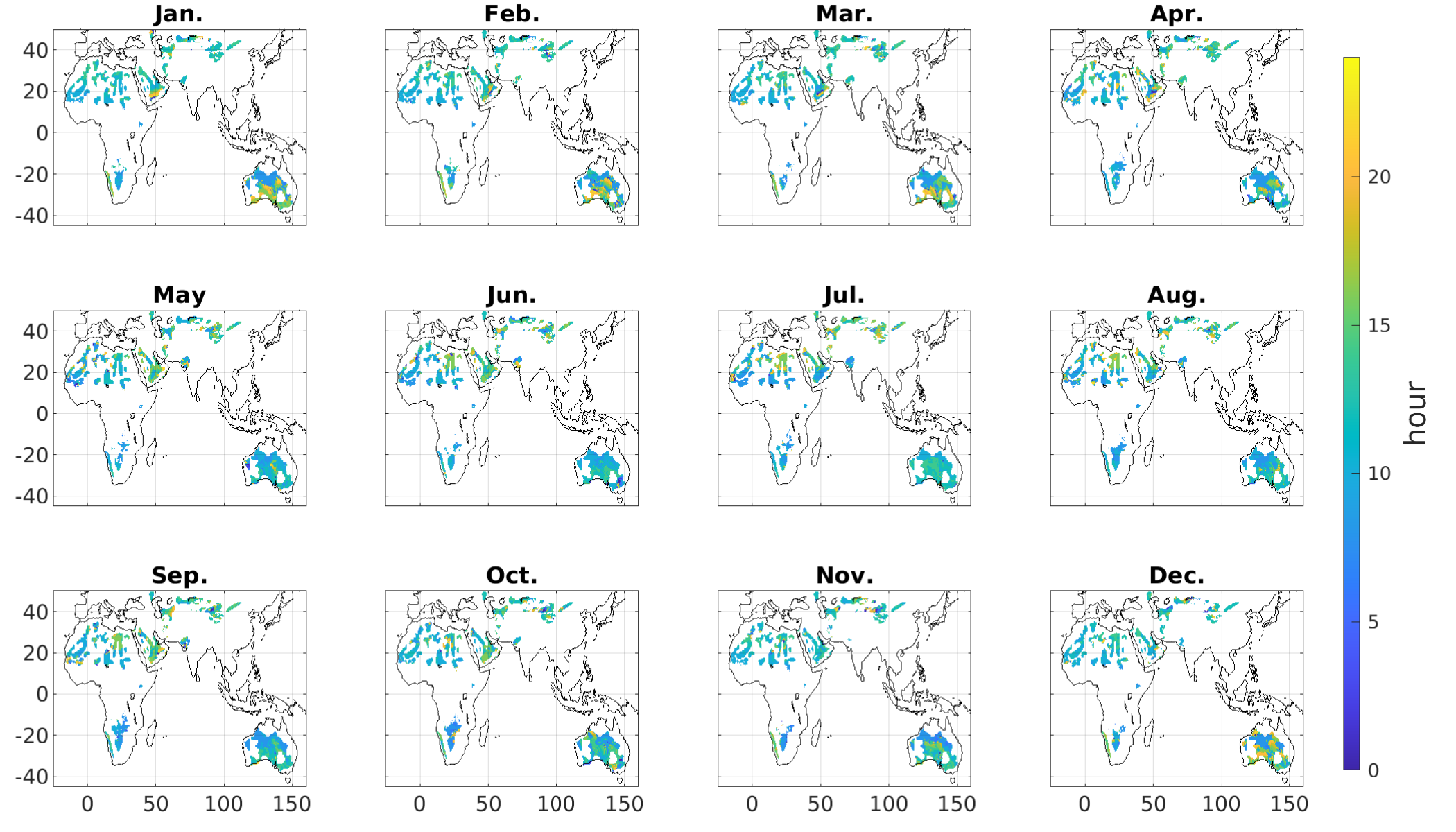}}
\caption{
  {\bf Time (hour) of maximum DP over dune regions during 2021 for different months.} 
\label{fig:time_max_dp_dunes}}
\end{figure}

\begin{figure}
%\centerline{\includegraphics[width=\linewidth]{./Figures/time_of_max_dp_year.pdf}}
\centerline{\includegraphics[width=\linewidth]{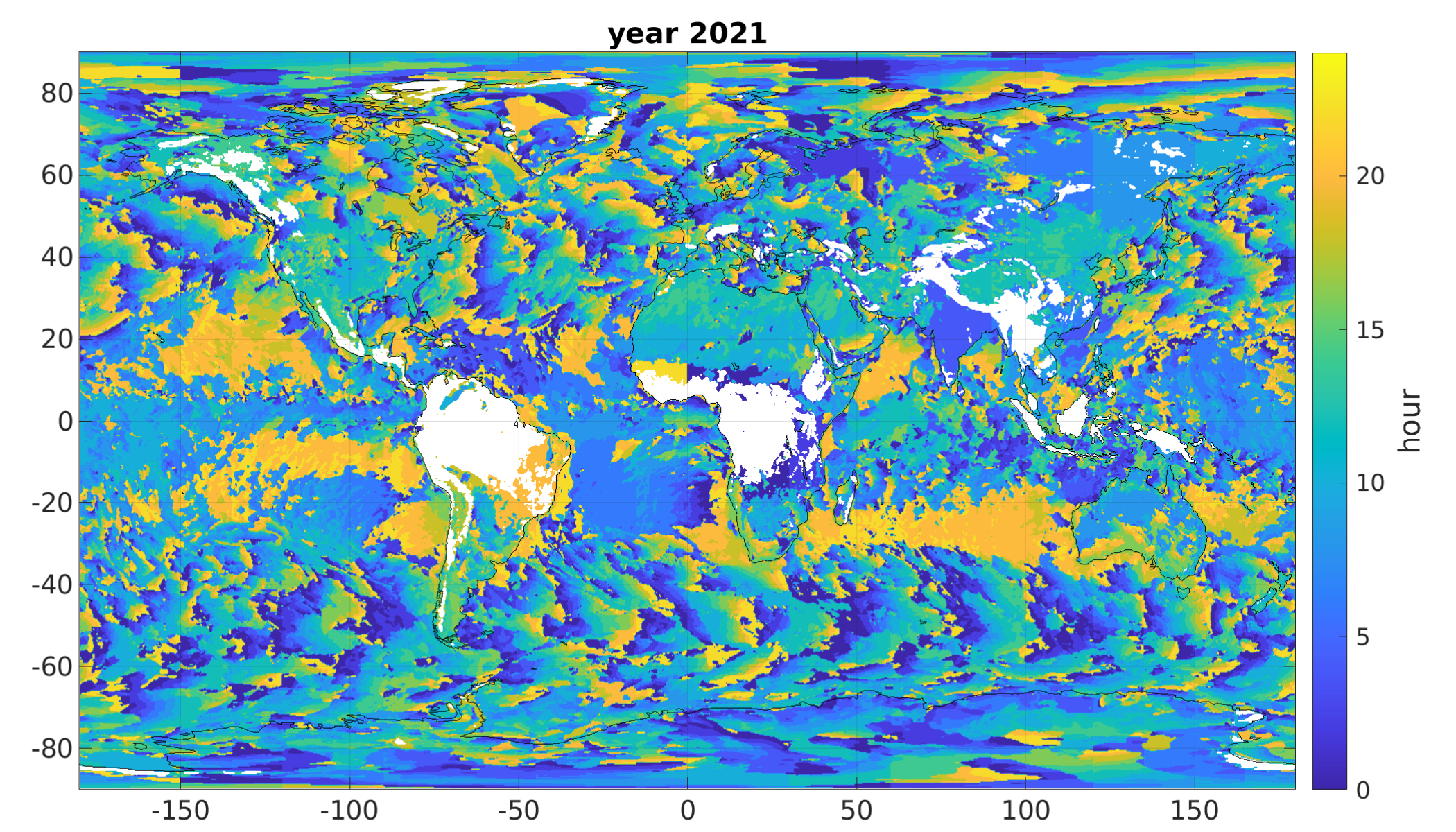}}
\caption{ {\bf Time (hour) of maximum DP during 2021.} The white color over the continental regions indicates regions with zero DP (i.e., regions over which the maximum velocity is smaller than the threshold velocity of $\sim$6 m s$^{-1}$).
\label{fig:time_max_dp_annual}}
\end{figure}

\begin{figure}
%\centerline{\includegraphics[width=\linewidth]{./Figures/time_of_max_dp_ocean.pdf}}
\centerline{\includegraphics[width=\linewidth]{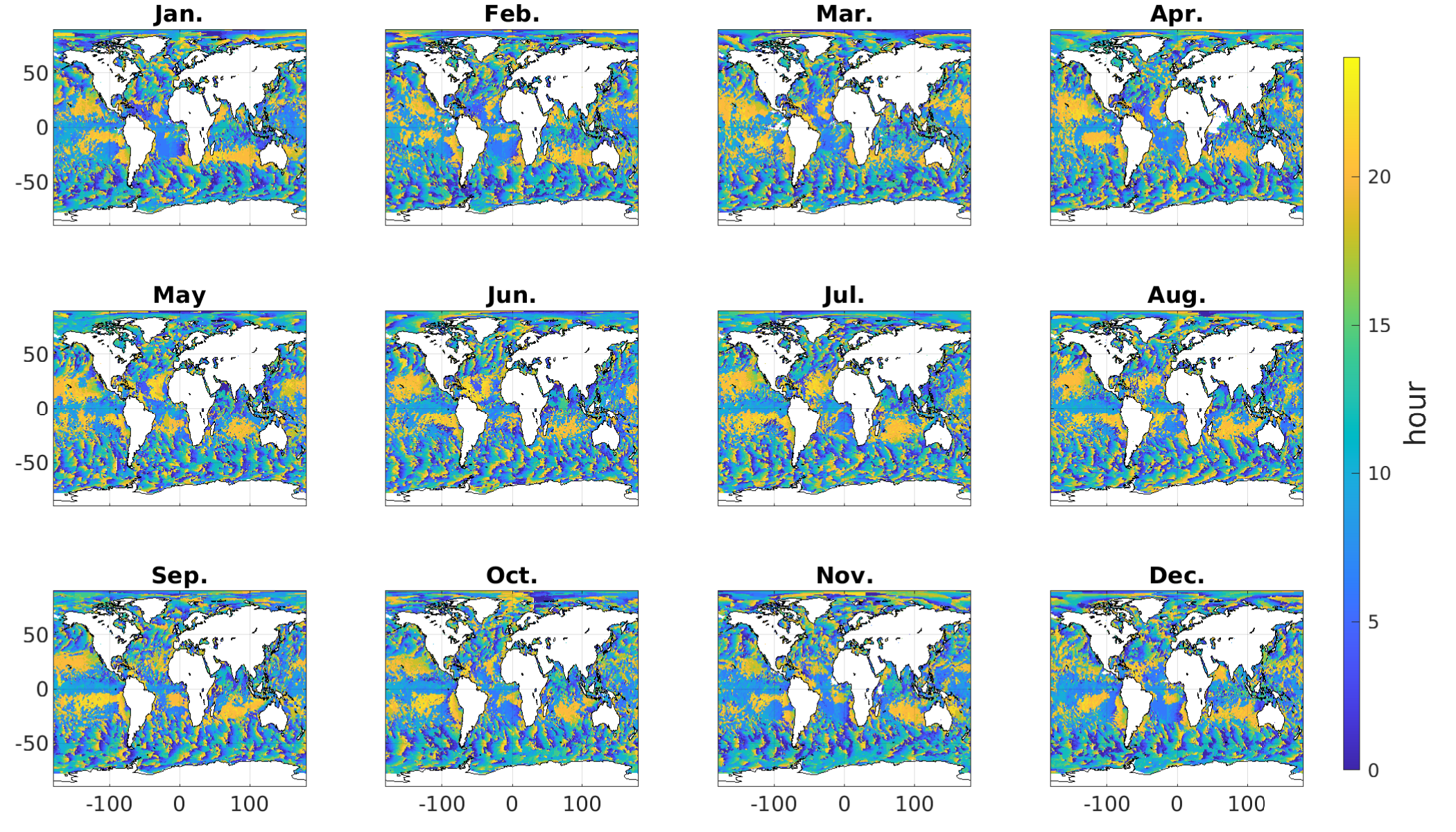}}
\caption{ {\bf Time of maximum DP over the ocean during 2021 for different months.}
\label{fig:time_max_dp_ocean}}
\end{figure}

\begin{figure}
%\centerline{\includegraphics[width=0.99\linewidth]{./Figures/time_of_max_dp_land.pdf}}
\centerline{\includegraphics[width=0.99\linewidth]{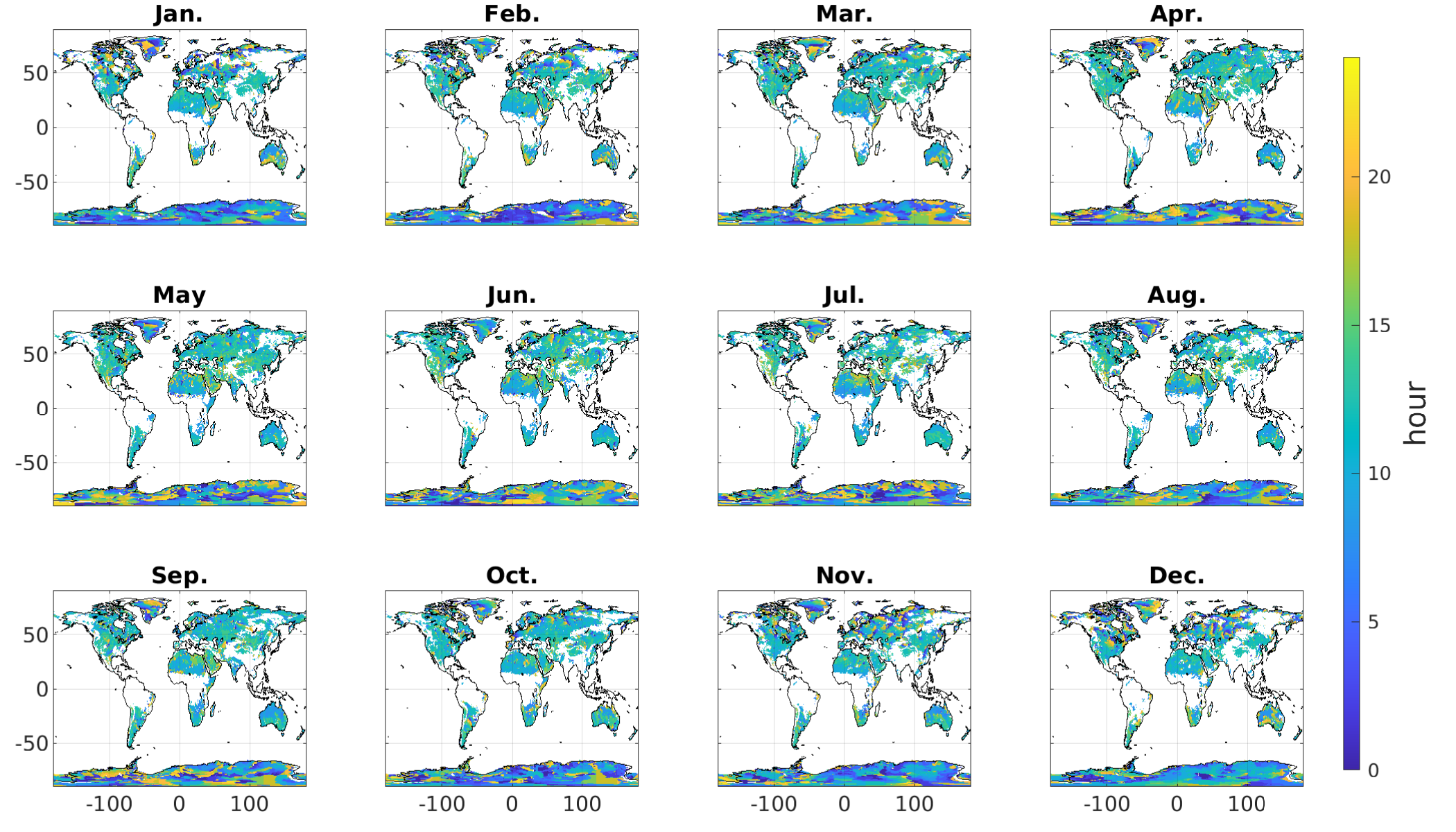}}
\caption{
  {\bf Time of maximum DP over the land during 2021 for different months.} The white color over the continental regions indicates regions with zero DP (i.e., regions over which the maximum velocity is smaller than the threshold velocity of $\sim$6 m s$^{-1}$).
\label{fig:time_max_dp_land}}
\end{figure}

\begin{figure}
%\centerline{\includegraphics[width=\linewidth]{./Figures/time_trends_spd.pdf}}
\centerline{\includegraphics[width=\linewidth]{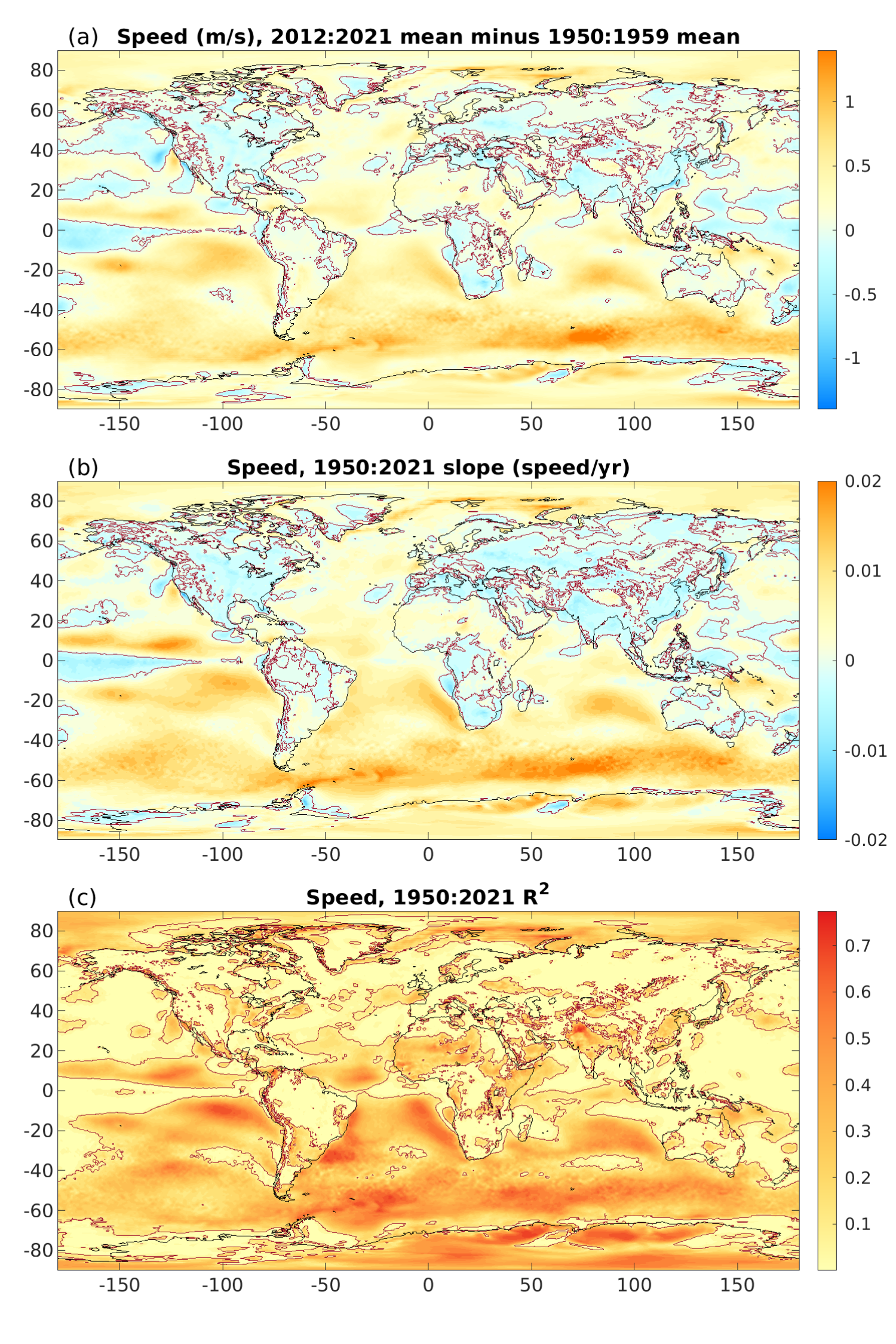}}
\caption{
  {\bf Wind speed analysis.} Same as Fig.~\ref{fig:time_trends_dp} for wind speed (m s$^{-1}$).  }
\label{fig:time_trends_spd}
\end{figure}

\begin{figure}
%\centerline{\includegraphics[width=\linewidth]{./Figures/time_trends_spd3.pdf}}
\centerline{\includegraphics[width=\linewidth]{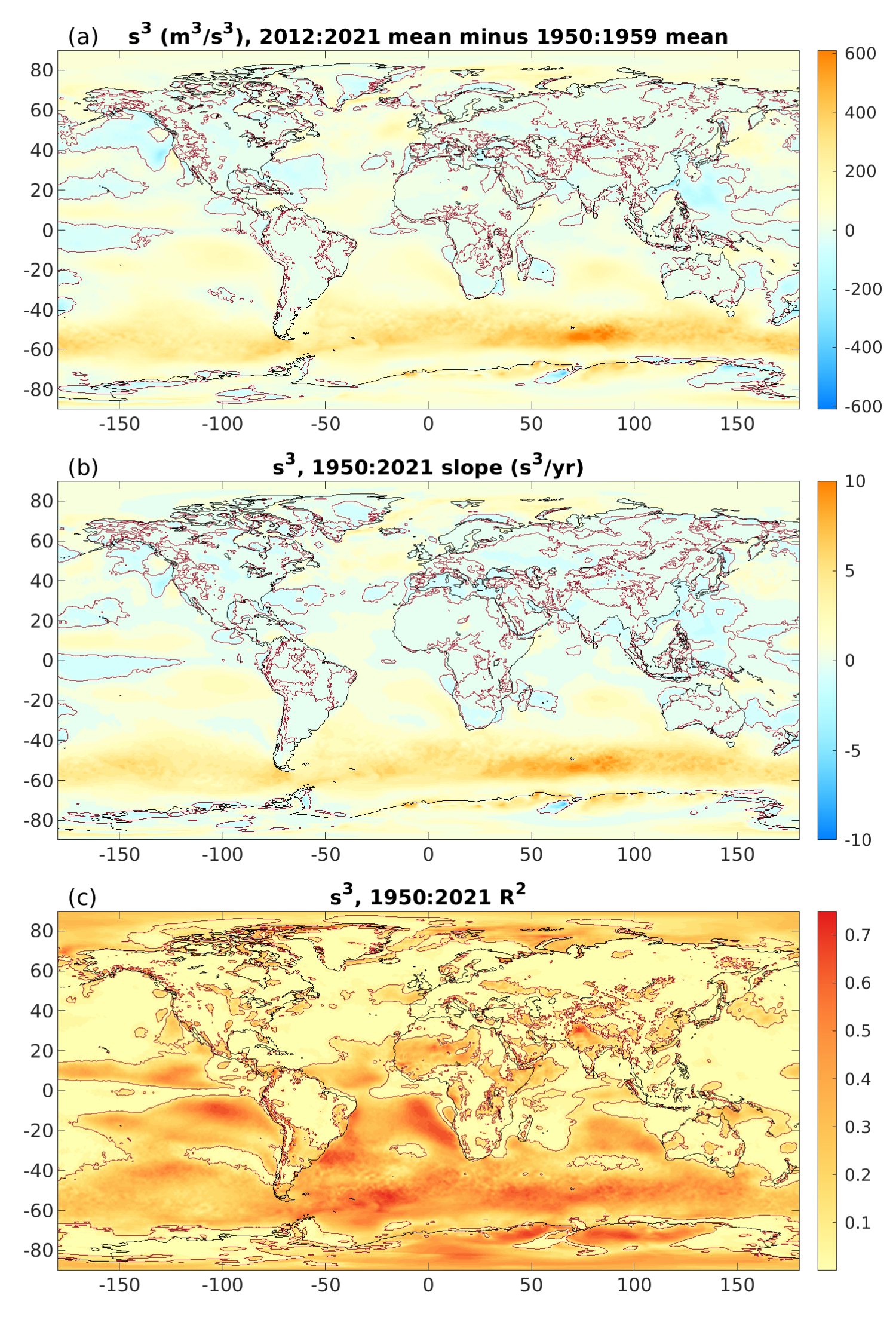}}
\caption{
  {\bf Wind power analysis.} Same as Fig.~\ref{fig:time_trends_dp} for the cube of the wind speed (m$^{3}$ s$^{-3}$) which is proportional to the wind power.   }
\label{fig:time_trends_spd3}
\end{figure}

\begin{figure}
%\centerline{\includegraphics[width=\linewidth]{./Figures/spd_time.pdf}}
\centerline{\includegraphics[width=\linewidth]{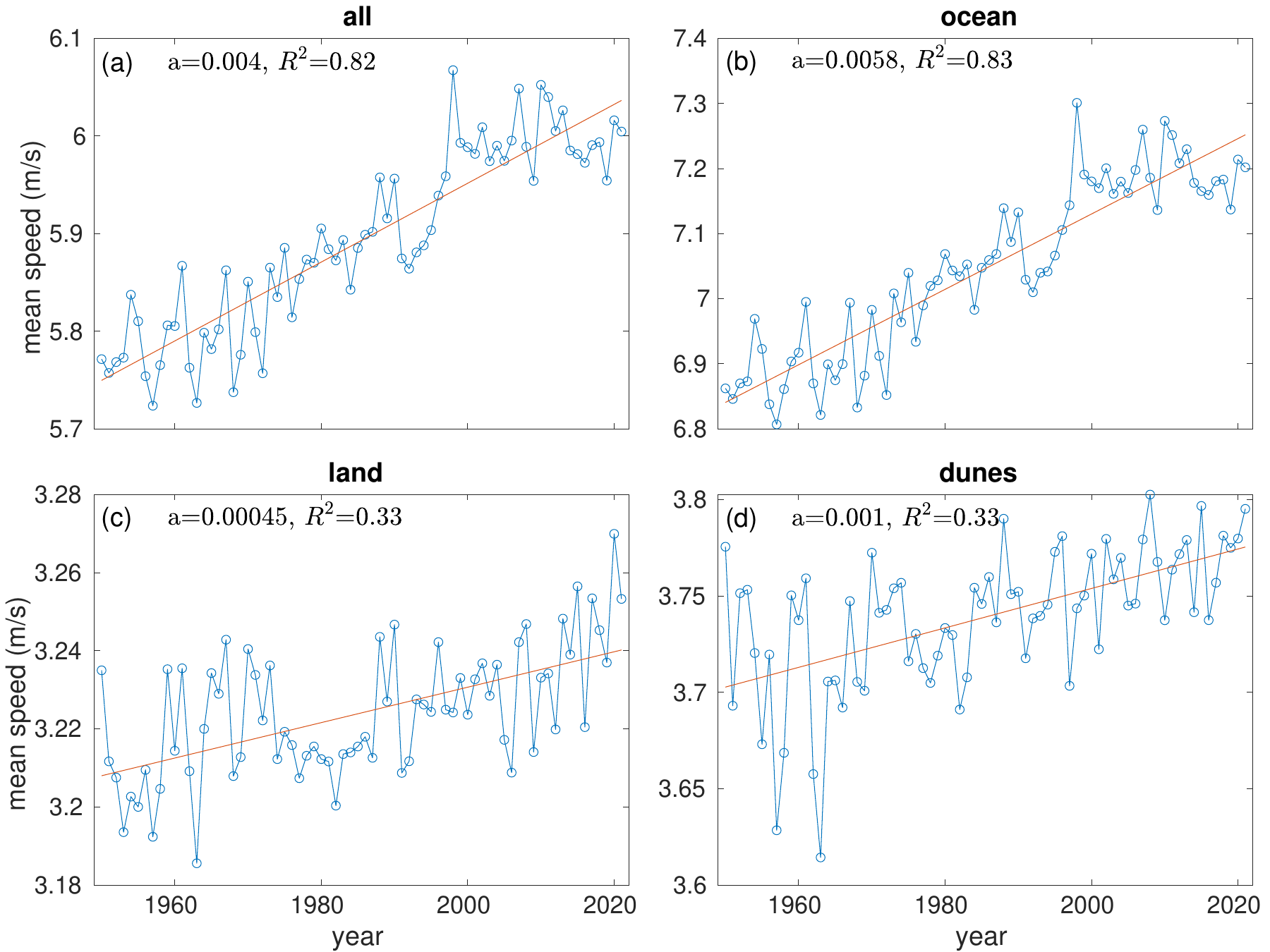}}
\caption{
  {\bf Wind speed trends.} Same as Fig.~\ref{fig:dp_time} for wind speed (m s$^{-1}$). }
\label{fig:spd_time}
\end{figure}

\begin{figure}
%\centerline{\includegraphics[width=\linewidth]{./Figures/era5_ncdc_us_eu_spd.pdf}}
\centerline{\includegraphics[width=\linewidth]{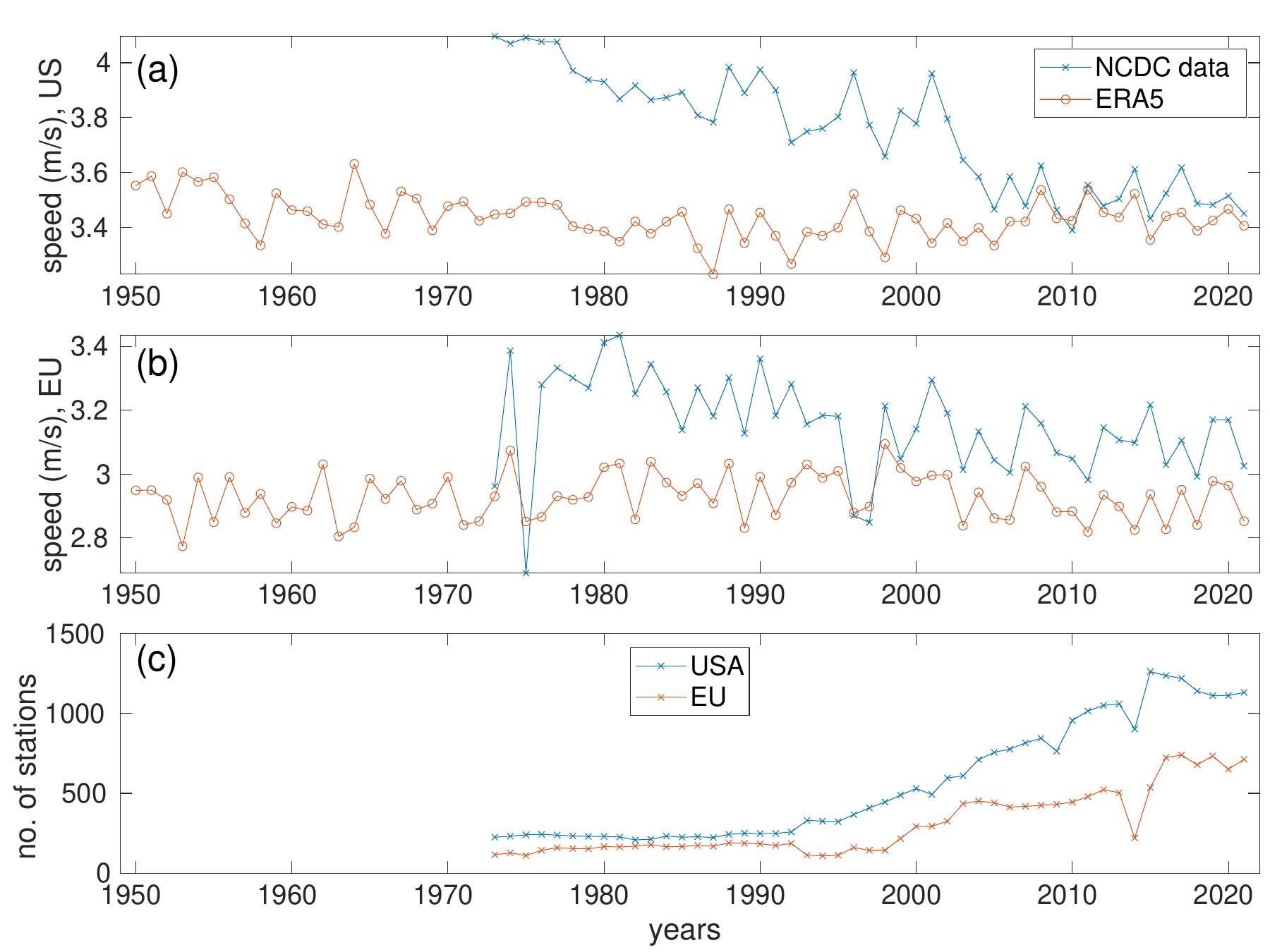}}
\caption{
  {\bf NCDC wind speed trends.}  (a) Same as Fig.~\ref{fig:era5_ncdc_us}c but for the USA wind speed. Note the better match between the two curves after 2007. The graph shows a decreasing trend in wind speed in the NCDC data but not in ERA5. (b) Same as panel a for the Europe wind speed. (c) The number of USA (blue) and EU (red) stations that were used in the analysis of this figure and of Fig.~\ref{fig:era5_ncdc_us}. }
\label{fig:era5_ncdc_us_spd}
\end{figure}

\begin{figure}
%\centerline{\includegraphics[width=\linewidth]{./Figures/SH_time.pdf}}
\centerline{\includegraphics[width=\linewidth]{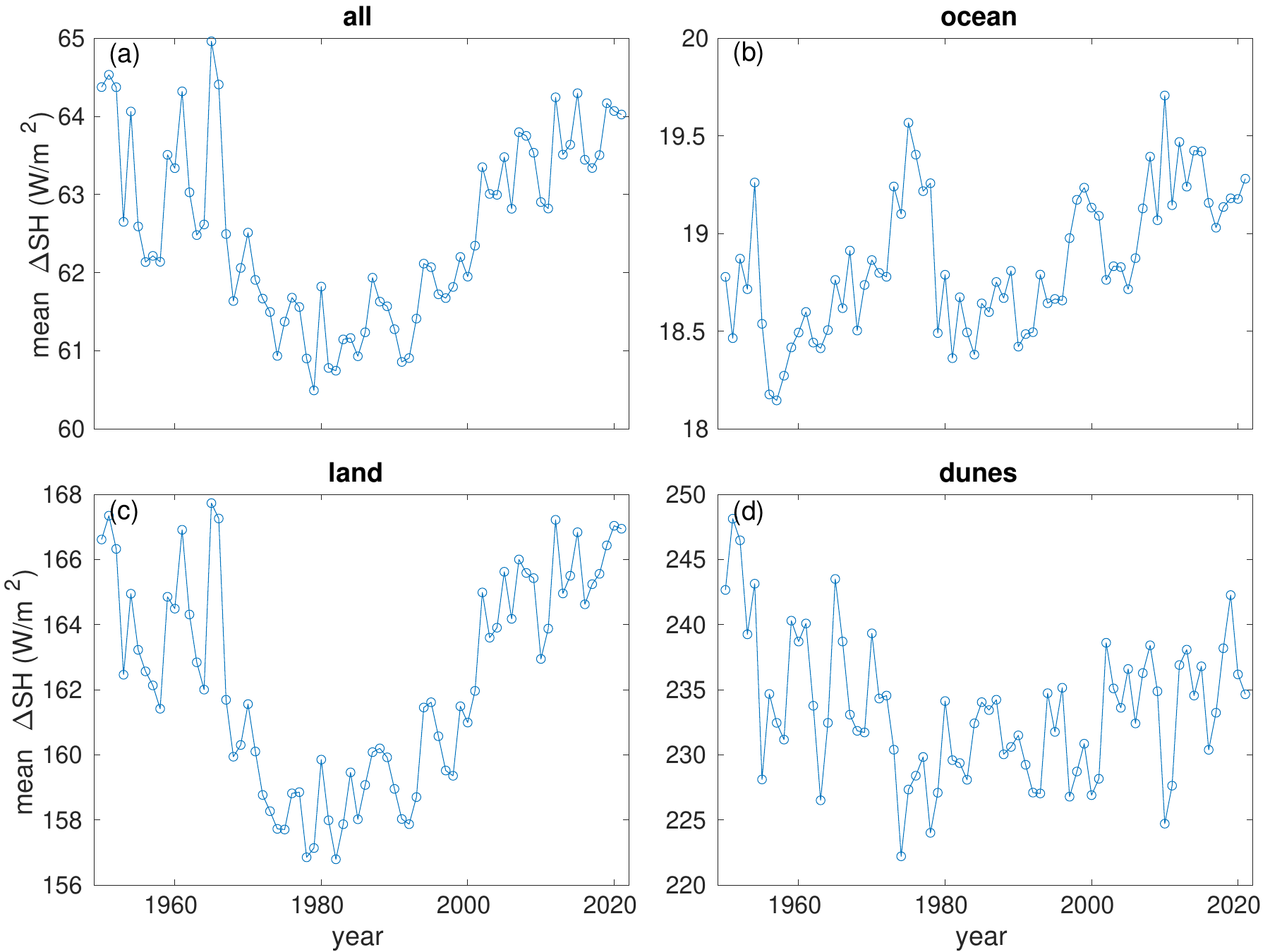}}
\caption{
  {\bf Sensible heat flux trends.} (a) The annual mean difference between the daily maximum and daily minimum sensible heat flux versus time over: (a) the entire globe, (b) the ocean, (c) the land, and (d) the dune regions. The largest values are over the dune regions as these regions experience the largest difference between day and night temperatures. }
\label{fig:SH_time}
\end{figure}

\end{document}